\documentclass[10pt,oneside,onecolumn,american,latin9]{aa}
\usepackage[T1]{fontenc}
\usepackage[latin9]{inputenc}
\usepackage[a4paper]{geometry}
\usepackage{babel}
\usepackage{array}
\usepackage{multirow}
\usepackage[fleqn]{amsmath}
\usepackage{graphicx}
\usepackage{wasysym}
\usepackage{esint}
\usepackage[unicode=true]
 {hyperref}
\usepackage{natbib}

\makeatletter

\providecommand{\tabularnewline}{\\}

\@ifundefined{date}{}{\date{}}
\makeatother

\begin{document}

\title{{\normalsize{}Cross-correlation of CFHTLenS galaxy catalogue and
Planck CMB lensing}\thanks{Based on observations obtained with Planck (\protect\href{http://www.esa.int/Planck}{http://www.esa.int/Planck}),
an ESA science mission with instruments and contributions directly
funded by ESA Member States, NASA, and Canada.}{\normalsize{} using the halo model prescription}}

\author{Adrien Kuntz \inst{1}}\institute{Physics department, \'{E}cole Normale Sup\'{e}rieure, 45 rue d'Ulm, 75005 Paris, France \email{adrien.kuntz@ens.fr}}

\abstract{}{I cross-correlate the galaxy counts from the Canada-France Hawaii
Telescope Lensing Survey (CFHTLenS) galaxy catalogue and cosmic microwave
background (CMB) convergence from the Planck data releases 1 (2013)
and 2 (2015).}{
I improve on an earlier study by computing an analytic covariance from the
halo model, implementing simulations to validate the theoretically
estimated error bars and the reconstruction method, fitting both a
galaxy bias and a cross-correlation amplitude using the joint cross
and galaxy auto-correlation, and performing a series of null tests.}{
Using a Bayesian analysis, I find a galaxy bias $b=0.92_{-0.02}^{+0.02}$
and a cross-correlation amplitude $A=0.85_{-0.16}^{+0.15}$ for the
2015 release, whereas for the 2013 release, I find $b=0.93_{-0.02}^{+0.02}$
and $A=1.05_{-0.15}^{+0.15}$.}{ I thus confirm the difference between
the two releases found earlier, although
both values of the amplitude now appear to be compatible with the
fiducial value $A=1$.}

\keywords{large-scale structure of the Universe - Cosmology : observations
- Cosmology : theory - Gravitational lensing : weak }\titlerunning{Cross-correlation of CFHTLenS catalogue and Planck CMB lensing}

\maketitle

\section{Introduction}

In the framework of the standard cosmological model, galaxies form
in matter overdensities that are the result of the non-linear growth
of primordial inhomogeneities generated by inflation. As a photon
travels from its surface of last scattering to us, its path is deflected
by the large-scale structures of the universe. Studying this weak
gravitational lensing of the cosmic microwave background (CMB), which is characterized
by temperature and polarization anisotropies, allows us to reconstruct
a map of the integrated (over the line of sight) overdensity of matter
of the universe \citep{okamoto2003cmblensing}.

Galaxies are expected to form inside dark matter halos, situated at
the peaks of the density fluctuations. Galaxies are therefore expected
to be good tracers of the large-scale structures, although their clustering
characteristics may be different from the dark matter ones. The ratio
between galaxy counts fluctuations and dark matter fluctuations is
called the galaxy bias. Studying the cross-correlation of lensing
convergence with galaxies allows this galaxy bias to be determined by
a method possibly free of unaccounted-for correlated systematics effects,
contrary to the auto-correlation of galaxies.

Several galaxy catalogues have been cross-correlated with the lensing
convergence: Planck CMB lensing cross-correlated with NVSS quasars,
MaxBCG clusters, SDSS LRGs, and the WISE Catalogue \citep{planckcollaboration2014textitplanck},
with the CFHTLenS galaxy catalogue \citep{omori2015crosscorrelation}
and with high-z submillimetre galaxies detected by the Herschel-ATLAS
survey \citep{bianchini2015crosscorrelation}; WMAP lensing cross-correlated
with NVSS galaxies \citep{smith2007detection} and with LRGs and quasars
from SDSS \citep{hirata2008correlation}; South Pole Telescope lensing
cross-correlated with Blanco Cosmology Survey galaxies \citep{bleem2012ameasurement}
and with WISE quasars \citep{geach2013adirect}; Atacama Cosmology
Telescope lensing cross-correlated with SDSS quasars \citep{sherwin2012theatacama}.

In this paper, I cross-correlate the galaxy counts from the Canada-France
Hawaii Telescope Lensing Survey (CFHTLenS) (\cite{heymans2012cfhtlens}; \cite{erben2013cfhtlens}) and the convergence all-sky map from
the Planck collaboration \citep{planckcollaboration2014textitplanck, planckcollaboration2015planck2015}
for the 2013 and 2015 releases. I followed the study of \cite{omori2015crosscorrelation},
who found the surprising result of a significant difference between
the galaxy bias inferred from the two releases. To complete their
work, I computed the theoretical covariances inferred from the halo
model \citep{cooray2002halomodels}, fitted the joint cross and auto-correlation,
implemented Gaussian simulations to check the error bars and the reconstruction
method, and performed a series of null tests. The difference between
the two releases, as found by \cite{omori2015crosscorrelation}, is recovered,
although it is less significant.

This paper is organized as follows. In Section \ref{sec:Theoretical-Background},
I present the theoretical background needed for this study and correct
some incomplete formulaes from the halo model. Data maps are presented
in Section \ref{sec:Data}, and the joint cross and auto-correlation
analysis is performed in Section \ref{sec:Constraints-on-Galaxy}.
Section \ref{sec:Consistency-Checks} presents the consistency checks
I carried out: Gaussian simulations and null tests. Finally
in Section \ref{sec:Summary-and-Conclusions} I summarize my results.

Throughout this paper, I assume a flat $\Lambda$CDM cosmology with
$h=0.70$, $H_{0}=100h\:\mathrm{km\,s^{-1}\,Mpc^{-1}}$, $\Omega_{\Lambda}^{0}=0.7$,
$\Omega_{m}^{0}=0.3$, $n_{s}=0.97$, $\sigma_{8}=0.82$, and $a_{0}=1$.

\section{Theoretical background\label{sec:Theoretical-Background}}

\subsection{Cross- and auto-correlation}

The effect of weak gravitational lensing by large-scale structures
on the CMB photons is described by a distortion matrix $\mathcal{\boldsymbol{\mathcal{A}}}$
that relates the direction of observation $\boldsymbol{\theta}=\left(\theta_{1},\theta_{2}\right)$
and the direction of the unlensed source $\boldsymbol{\theta_{s}}$: $\mathbf{\boldsymbol{\theta}_{s}=\boldsymbol{\mathcal{A}\cdot\theta}}$
with $\mathcal{A}_{ab}=\delta_{ab}-\partial_{\theta_{a}\theta_{b}}\psi\left(\boldsymbol{\theta}\right)$. Here,
$\psi$ is the lensing potential given by e.g. \cite{peter2013primordial},
pp 398-399:
\begin{equation}
\psi\left(\boldsymbol{\theta}\right)=\frac{2}{c^{2}}\int_{0}^{\chi_{CMB}}\frac{\chi_{CMB}-\chi}{\chi\chi_{CMB}}\Phi\left[\chi\boldsymbol{\theta},\chi\right]d\chi
.\end{equation}

In this equation $\chi$ is the line-of-sight comoving distance, $\chi_{CMB}$
is the comoving distance of the CMB at redshift $z_{CMB}\simeq1090$,
and $\Phi$ is the gravitational potential at the point on the photon
path given by $\chi\boldsymbol{\theta}$. The lensing convergence
$\kappa$ is defined as $\kappa\left(\boldsymbol{\theta}\right)\equiv\triangle_{2}\psi\left(\boldsymbol{\theta}\right)/2$
and is related to the matter overdensity $\delta$ via  
\begin{equation}
\kappa\left(\boldsymbol{\theta}\right)=\intop_{0}^{\chi_{CMB}}W^{\kappa}\left(\chi\right)\delta\left(\chi\boldsymbol{\theta},\chi\right)d\chi,\label{eq:kappa(theta)}
\end{equation}
\begin{equation}
W^{\kappa}\left(\chi\right)=\frac{3}{2}\Omega_{m}^{0}\left(\frac{H_{0}}{c}\right)^{2}\frac{\chi\left(\chi_{CMB}-\chi\right)}{\chi_{CMB}}\left(1+z\left(\chi\right)\right)
.\end{equation}

\begin{figure}[h]
\includegraphics[width=0.8\paperwidth]{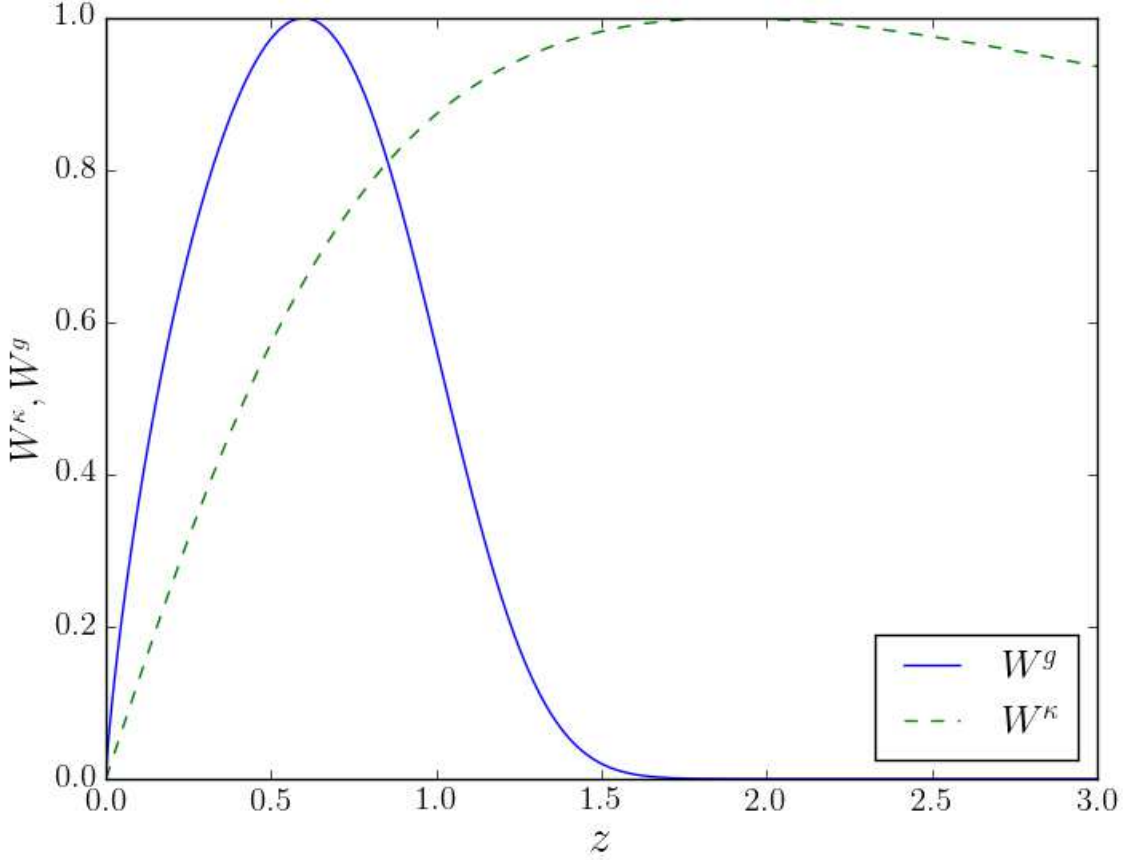}

\protect\caption{Lensing kernel $W^{\kappa}$ (dashed line) and galaxy overdensity
kernel $W^{g}$ (solid line) for all patches of the CFHTLenS catalogue.
Both kernels are multiplied by $d\chi/dz$ and normalized to a unit
maximum.\label{fig:WkWg}}

\end{figure}

The overdensity of galaxies is defined as 
\begin{equation}
g\left(\boldsymbol{\theta}\right)=\intop_{0}^{\chi_{CMB}}W^{g}\left(\chi\right)\delta\left(\chi\boldsymbol{\theta},\chi\right)d\chi,
\end{equation}
\begin{equation}
W^{g}\left(\chi\right)=\frac{dN\left(z\right)}{dz}\frac{dz}{d\chi}b\left(\chi\right)+\frac{3}{2}\Omega_{m}^{0}\left(\frac{H_{0}}{c}\right)^{2}\left(1+z\left(\chi\right)\right)\left(5s-2\right)f\left(\chi\right)\label{eq:Wg}
\end{equation}
with
\begin{equation}
f\left(\chi\right)=\chi\intop_{\chi}^{\chi_{CMB}}d\chi'\frac{\chi'-\chi}{\chi'}\frac{dN}{dz}\frac{dz}{d\chi'}
\end{equation}
and
\[
s=\frac{d\log_{10}N\left(<m\right)}{dm}.
\]
The second term in Equation (\ref{eq:Wg}) is the magnification bias
\citep{moessner1998theeffect}, which occurs because the
number density of galaxies is altered by gravitational lensing; it has an effect of a few percent. The $dN/dz$ ratio is the redshift
distribution of the galaxy sample normalized such that $\intop dz\left(dN/dz\right)=1$.
The over-density of galaxies is assumed to be linearly proportional
to the matter over-density: $\delta^{g}\left(\chi\boldsymbol{\theta},\chi\right)=b\left(\chi\right)\delta\left(\chi\boldsymbol{\theta},\chi\right)$.
In this article, the linear bias is assumed to be independent of $\chi$,
which is a rather good approximation given the sharply peaked redshift
distribution (see Figure \ref{fig:Redshift-distribution}). The fiducial
value adopted for the galaxy bias is $b=1,$ as is suggested in the study
of \cite{omori2015crosscorrelation}. Because the effect of the magnification
bias is very weak, I take $b$ as an overall factor of $W^{g}$ for
simplicity in the fitting algorithm. Figure \ref{fig:WkWg} shows
a plot of the two kernels $W^{\kappa}$ and $W^{g}$.

Decomposing $\kappa\left(\boldsymbol{\theta}\right)\equiv\sum_{lm}\kappa_{lm}Y_{l}^{m}\left(\boldsymbol{\theta}\right)$
and $g\left(\boldsymbol{\theta}\right)\equiv\sum_{lm}g_{lm}Y_{l}^{m}\left(\boldsymbol{\theta}\right)$
into spherical harmonics and using the Limber approximation \citep{loverde2008extended}
for small angles, which consists in approximating the spherical Bessel
functions by $j_{l}\left(x\right)=\sqrt{\pi/\left(2l+1\right)}\delta_{Dirac}\left(x-l-1/2\right)$
yields 
\begin{equation}
C_{l}^{\kappa g}=\intop d\chi\frac{W^{\kappa}\left(\chi\right)W^{g}\left(\chi\right)}{\chi^{2}}P\left(k=\frac{l}{\chi},\,z\left(\chi\right)\right)\label{eq:Clkg}
,\end{equation}

\noindent where $C_{l}^{\kappa g}$, the cross-correlation between $\kappa$
and $g$, is defined as $\left\langle \kappa_{lm}g_{l'm'}^{*}\right\rangle =\delta_{ll'}\delta_{mm'}C_{l}^{\kappa g}$,
and $P\left(k,\,z\right)$ is the matter power spectrum, which I compute
using the halo model (see Section \ref{sec:The-halo-model}) under
the convention that 
\begin{eqnarray}
\delta\left(\chi\boldsymbol{\theta},\chi\right) & \equiv & \intop\frac{d^{3}\boldsymbol{k}}{\left(2\pi\right)^{3}}e^{-i\boldsymbol{k}\cdot\left(\chi\boldsymbol{\theta},\chi\right)}\delta\left(\boldsymbol{k}\right)\\
\left\langle \delta\left(\boldsymbol{k}\right)\delta\left(\boldsymbol{k'}\right)\right\rangle  & \equiv & \left(2\pi\right)^{3}\delta_{D}\left(\boldsymbol{k}+\boldsymbol{k'}\right)P\left(k\right)\\
\left\langle \delta\left(\boldsymbol{k_{1}}\right)\delta\left(\boldsymbol{k_{2}}\right)\delta\left(\boldsymbol{k_{3}}\right)\right\rangle  & \equiv & \left(2\pi\right)^{3}\delta_{D}\left(\boldsymbol{k_{1}}+\boldsymbol{k_{2}}+\boldsymbol{k_{3}}\right)B\left(\boldsymbol{k_{1}},\,\boldsymbol{k_{2}},\,\boldsymbol{k_{3}}\right)\\
\left\langle \delta\left(\boldsymbol{k_{1}}\right)\delta\left(\boldsymbol{k_{2}}\right)\delta\left(\boldsymbol{k_{3}}\right)\delta\left(\boldsymbol{k_{4}}\right)\right\rangle  & \equiv & \left(2\pi\right)^{3}\delta_{D}\left(\boldsymbol{k_{1}}+\boldsymbol{k_{2}}+\boldsymbol{k_{3}}+\boldsymbol{k_{4}}\right)T\left(\boldsymbol{k_{1}},\,\boldsymbol{k_{2}},\,\boldsymbol{k_{3}},\,\boldsymbol{k_{4}}\right)\\
 & ...\nonumber 
\end{eqnarray}
where $\delta_{D}$ stands for the Dirac delta function.

The auto-spectra are computed the same way:

\begin{equation}
C_{l}^{\kappa\kappa}=\intop d\chi\frac{W^{\kappa}\left(\chi\right)^{2}}{\chi^{2}}P\left(k=\frac{l}{\chi},\,z\left(\chi\right)\right)\label{eq:Clkk}
\end{equation}

\begin{equation}
C_{l}^{gg}=\intop d\chi\frac{W^{g}\left(\chi\right)^{2}}{\chi^{2}}P\left(k=\frac{l}{\chi},\,z\left(\chi\right)\right)\label{eq:Clgg}
.\end{equation}

\noindent I then need an estimator for the cross- and auto-spectra and its covariance, and this is the subject of the next section.

\subsection{Estimator}

The estimator used for the cross-correlation of the datasets is
\begin{equation}
\tilde{C}_{l_{i}}^{\kappa g}=\frac{1}{f_{sky}}\frac{1}{\left|B_{i}\right|}\sum_{l\in B_{i}}\frac{1}{2l+1}\sum_{m=-l}^{l}\tilde{\kappa}_{lm}\tilde{g}_{lm}^{*}
,\end{equation}
replacing $g$ with $\kappa$ for the autocorrelations. Quantities
with a tilde are observed data, $f_{sky}$ is the fraction of sky
covered by the datasets, $B_{i}$ is the bin in $l$ used for the
estimator, which is taken in this study as ranging from $l_{min}=50$
to $l_{max}=1950$ with width $\delta l=100$, which corresponds to a number
of bins $N_{bin}=19$. The lower $l$ cut is here because of the limited
coverage of the sky imposed by the galaxy catalogue, the correlation
being meaningful only below a few degrees. I numerically compute the
covariance of this estimator using the halo model. Without any other
approximation, the full covariance would be too heavy to compute, since it
would involve six integrations. That is why I use the flat-sky approximation
\citep{bernardeau2010cmbspectra}, which consists in approximating
the sphere by its tangential plane, and so is only valid at small
angles. The spherical harmonics transform is then replaced by a simple
Fourier transform:
\begin{equation}
\kappa\left(\boldsymbol{l}\right)\equiv\frac{1}{\sqrt{4\pi}}\intop d^{2}\boldsymbol{\theta}e^{i\boldsymbol{l}\cdot\boldsymbol{\theta}}\kappa\left(\boldsymbol{\theta}\right)
.\end{equation}

The normalization is here to ensure that $\kappa\left(\boldsymbol{0}\right)=\kappa_{00}$.
The same equation applies for $g$. Using Equation (\ref{eq:kappa(theta)}),
one finds
\begin{equation}
\kappa\left(\boldsymbol{l}\right)=\frac{1}{\sqrt{4\pi}}\intop_{0}^{\chi_{CMB}}d\chi\frac{W^{\kappa}\left(\chi\right)}{\chi^{2}}\intop_{-\infty}^{+\infty}\frac{dk}{2\pi}\delta\left(\frac{\boldsymbol{l}}{\chi},\,k\right)e^{-ik\chi}
,\end{equation}
and the correlator is
\begin{equation}
\left\langle \kappa\left(\boldsymbol{l}\right)g^{*}\left(\boldsymbol{l'}\right)\right\rangle =\frac{1}{4\pi}\intop d\chi d\chi'\frac{W^{\kappa}\left(\chi\right)}{\chi^{2}}\frac{W^{g}\left(\chi'\right)}{\chi'^{2}}\left(2\pi\right)^{2}\delta_{D}\left(\frac{\boldsymbol{l}}{\chi}+\frac{\boldsymbol{l'}}{\chi'}\right)\intop_{-\infty}^{+\infty}\frac{dk}{2\pi}e^{-ik\left(\chi-\chi'\right)}P\left(\sqrt{\frac{l^{2}}{\chi^{2}}+k^{2}}\right)
.\end{equation}

In the small angle approximation, $k$ is neglected before $l/\chi$.
(For high values of $k$ the integral is suppressed because of the
oscillatory function.) This yields 
\begin{equation}
\left\langle \kappa\left(\boldsymbol{l}\right)g\left(\boldsymbol{l'}\right)\right\rangle =\pi\delta_{D}\left(\boldsymbol{l}+\boldsymbol{l'}\right)C_{l}^{\kappa g}
.\end{equation}

\noindent The estimator of $C_{l}^{\kappa g}$ in the flat-sky approximation
is
\begin{equation}
\tilde{C}_{l_{i}}^{\kappa g}\equiv\frac{1}{f_{sky}}\intop_{\left|l\right|\in B_{i}}\frac{d^{2}\boldsymbol{l}}{\Omega_{i}}\tilde{\kappa}_{\mathcal{W}}\left(\boldsymbol{l}\right)\tilde{g}_{\mathcal{W}}\left(-\boldsymbol{l}\right)
.\end{equation}

\noindent The subscript $\mathcal{W}$ refers to the mask function:\ Because the
different surveys only probe a part of the sky, what is measured is
$\tilde{\kappa}_{\mathcal{W}}\left(\boldsymbol{\theta}\right)\equiv\mathcal{W}_{\kappa}\left(\boldsymbol{\theta}\right)\tilde{\kappa}\left(\boldsymbol{\theta}\right)$
and $\tilde{g}_{\mathcal{W}}\left(\boldsymbol{\theta}\right)\equiv\mathcal{W}_{g}\left(\boldsymbol{\theta}\right)\tilde{g}\left(\boldsymbol{\theta}\right)$
with $\mathcal{W}=1$ where the data are not masked and $\mathcal{W}=0$
where the data are masked. Here, $\Omega_{i}\equiv\pi\left(\left(l_{i}+\delta l_{i}\right)^{2}-l_{i}^{2}\right)\simeq2\pi l_{i}\delta l_{i}$
is the size of the bin $i$, and $f_{sky}\equiv\intop d^{2}\boldsymbol{\theta}\mathcal{W}_{\kappa}\left(\boldsymbol{\theta}\right)\mathcal{W}_{g}\left(\boldsymbol{\theta}\right)/\left(4\pi\right)$
is here to ensure that $\left\langle \tilde{C}_{l_{i}}^{\kappa g}\right\rangle =C_{l_{i}}^{\kappa g}$
.Indeed,
\begin{eqnarray}
\left\langle \tilde{C}_{l_{i}}^{\kappa g}\right\rangle  & = & \frac{1}{f_{sky}}\intop_{\left|l\right|\in B_{i}}\frac{d^{2}\boldsymbol{l}}{\Omega_{i}}\frac{1}{4\pi}\intop\frac{d^{2}\boldsymbol{l'}}{\left(2\pi\right)^{2}}C_{\left|\boldsymbol{l}-\boldsymbol{l'}\right|}\tilde{\mathcal{W}}_{\kappa}\left(\boldsymbol{l'}\right)\tilde{\mathcal{W}}_{g}\left(-\boldsymbol{l'}\right)\nonumber \\
 & \simeq & \frac{C_{l_{i}}^{\kappa g}}{f_{sky}}\frac{1}{4\pi}\intop\frac{d^{2}\boldsymbol{l'}}{\left(2\pi\right)^{2}}\tilde{\mathcal{W}}_{\kappa}\left(\boldsymbol{l'}\right)\tilde{\mathcal{W}}_{g}\left(-\boldsymbol{l'}\right)\\
 & = & C_{l_{i}}^{\kappa g}
.\end{eqnarray}

\noindent To find this expression of $f_{sky}$ one must assume that the size
of the bin $\delta l_{i}$ is larger than the typical length of variation
in the Fourier transform of the mask functions, which is true in the
case studied here. ($\delta l=100$ and the typical size of a field
of galaxies is $\sim5$ degrees.)

The calculations are exactly the same for the autocorrelation, except
that one must pay attention to the noise in the
data, which are uncorrelated between the two maps:
\begin{eqnarray}
\left\langle \tilde{C}_{l_{i}}^{\kappa\kappa}\right\rangle  & = & C_{l_{i}}^{\kappa\kappa}+N_{l_{i}}^{\kappa\kappa}\\
\left\langle \tilde{C}_{l_{i}}^{gg}\right\rangle  & = & C_{l_{i}}^{gg}+N_{l_{i}}^{gg}
.\end{eqnarray}

The covariance of this estimator $\Sigma_{ij}^{\kappa g}=\left\langle \tilde{C}_{l_{i}}^{\kappa g}\tilde{C}_{l_{j}}^{\kappa g}\right\rangle -\left\langle \tilde{C}_{l_{i}}^{\kappa g}\right\rangle \left\langle \tilde{C}_{l_{j}}^{\kappa g}\right\rangle $
is calculated in the same way:
\begin{eqnarray}
\Sigma_{ij}^{\kappa g} & = & \frac{1}{4\pi f_{sky}}\left[\frac{\left(2\pi\right)^{2}}{\Omega_{i}}\delta_{ij}^{K}\left(\left(C_{l_{i}}^{\kappa\kappa}+N_{l_{i}}^{\kappa\kappa}\right)\left(C_{l_{j}}^{gg}+N_{l_{i}}^{gg}\right)+\left(C_{l_{i}}^{\kappa g}\right)^{2}\right)+\bar{T}_{ij}^{\kappa g}\right]\label{eq:variance}\\
\bar{T}_{ij}^{\kappa g} & = & \intop\frac{d^{2}\boldsymbol{l}}{\Omega_{i}}\frac{d^{2}\boldsymbol{l'}}{\Omega_{j}}\intop d\chi\frac{W^{\kappa}\left(\chi\right)^{2}W^{g}\left(\chi\right)^{2}}{\chi^{6}}T\left(\frac{\boldsymbol{l}}{\chi},\,-\frac{\boldsymbol{l}}{\chi},\,\frac{\boldsymbol{l'}}{\chi},\,-\frac{\boldsymbol{l'}}{\chi}\right)\label{eq:Tbarij}
.\end{eqnarray}

\begin{figure}[h]
\includegraphics[width=0.8\paperwidth]{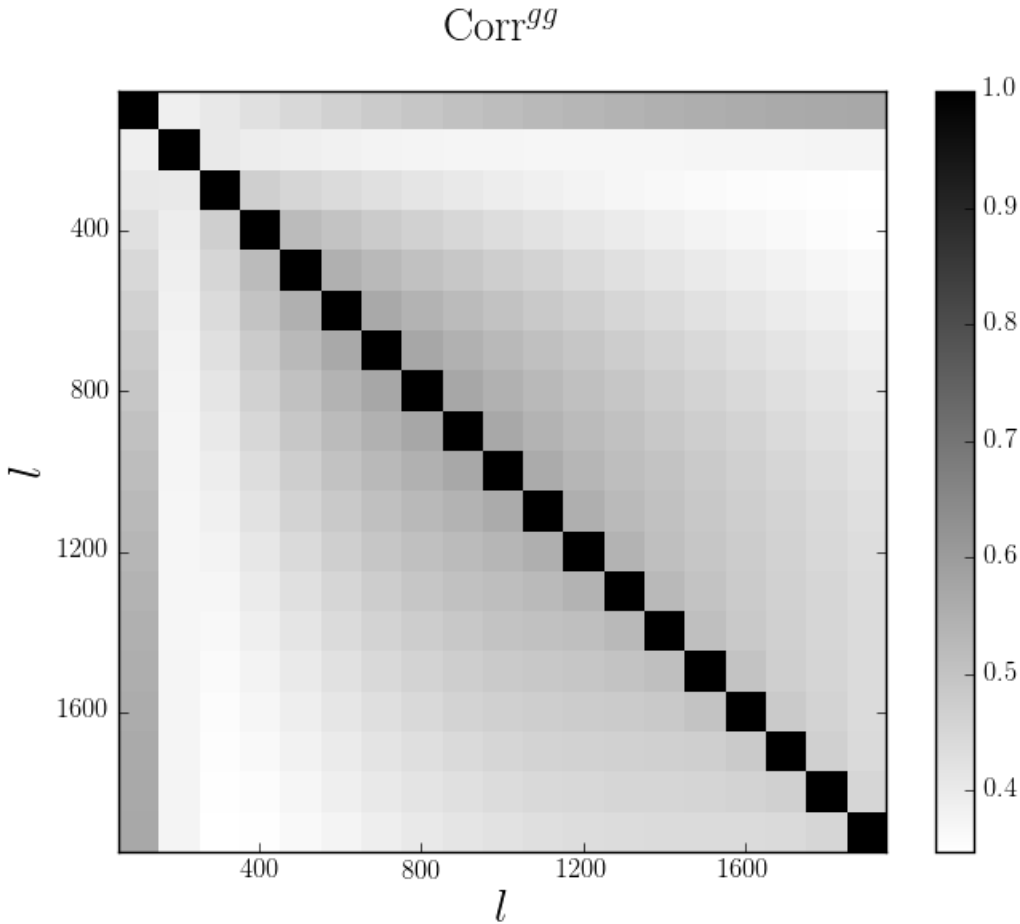}\protect\caption{Correlation matrix $\mathrm{Corr}^{gg}$ built with the covariance
of Equation (\ref{eq:variance}) for the galaxy autocorrelation, using
a binning of width $\delta l=100$ from $l_{min}=50$ to $l_{max}=1950$.
The non-Gaussian term has a strong amplitude, contrary to the one
in $\mathrm{Corr}^{\kappa\kappa}$ and $\mathrm{Corr}^{\kappa g}$
because of the very high noise reconstruction of the convergence $N_{l}^{\kappa\kappa}$.
\label{fig:Corgg}}

\end{figure}

In this computation, I have ignored the beat-coupling effect that may
arise from finite-volume survey effects \citep{takada2013powerspectrum}.
The first term in Equation (\ref{eq:variance}) is the Gaussian variance,
and the second term arises because of non-Gaussianity. It is important
to note that while both terms are inversely proportional to the volume
of the survey, only the non-Gaussian term remains constant with the
binning adopted. Therefore this term can gain significant importance
when the binning is large. The term $\delta_{ij}^{K}$ is the Kronecker
delta. The covariances for the autocorrelations read as
\begin{eqnarray}
\Sigma_{ij}^{\kappa\kappa} & = & \frac{1}{4\pi f_{sky}}\left[\frac{\left(2\pi\right)^{2}}{\Omega_{i}}2\delta_{ij}^{K}\left(C_{l_{i}}^{\kappa\kappa}+N_{l_{i}}^{\kappa\kappa}\right)^{2}+\bar{T}_{ij}^{\kappa\kappa}\right]\\
\bar{T}_{ij}^{\kappa\kappa} & = & \intop\frac{d^{2}\boldsymbol{l}}{\Omega_{i}}\frac{d^{2}\boldsymbol{l'}}{\Omega_{j}}\intop d\chi\frac{W^{\kappa}\left(\chi\right)^{4}}{\chi^{6}}T\left(\frac{\boldsymbol{l}}{\chi},\,-\frac{\boldsymbol{l}}{\chi},\,\frac{\boldsymbol{l'}}{\chi},\,-\frac{\boldsymbol{l'}}{\chi}\right)
,\end{eqnarray}
and similarly for the galaxies replacing $\kappa$ with $g$. For
the purpose of this study (cf Section \ref{sec:Constraints-on-Galaxy}),
I also need a mixed covariance defined as
\begin{equation}
\Sigma_{ij}^{\kappa g-gg}=\left\langle \tilde{C}_{l_{i}}^{\kappa g}\tilde{C}_{l_{j}}^{gg}\right\rangle -\left\langle \tilde{C}_{l_{i}}^{\kappa g}\right\rangle \left\langle \tilde{C}_{l_{j}}^{gg}\right\rangle \label{eq:sigmakg-gg}
\end{equation}
and computed using the same prescription as above. The computed correlation
matrix for the galaxy autocorrelation is shown in Figure \ref{fig:Corgg}.
It is clear that the non-Gaussian term has a non-negligible amplitude.
However, in the cross-correlation covariance $\Sigma^{\kappa g}$,
the noise reconstruction of the convergence map is very high, so
the Gaussian term dominates the non-Gaussian one.

\subsection{The halo model\label{sec:The-halo-model}}

The halo model is based on the spherical collapse model \citep{gunn1972onthe}: large-scale structures formed from sufficiently overdense regions
of space that collapsed under their own gravity. The remaining structures
are called halos. Fitting formulaes for the number density of halos
are presented in Section \ref{sub:The-Number-Density}. Section \ref{sub:Halo-Biasing}
is about halo biasing with respect to dark matter; Section \ref{sub:Halo-Profiles}
presents the profiles of halos; and finally in Section \ref{sub:The-Power-Spectrum},
I compute the power spectrum and trispectrum in the halo model. For
a thorough analysis of the halo model, see \cite{cooray2002halomodels}.

\subsubsection{The number density of halos\label{sub:The-Number-Density}}

A useful formula for the number density of halos at redshift $z$,
$n\left(m,z\right)$ is provided by \cite{sheth1999largescale}, following
an original formula by \cite{press1974formation}:
\begin{equation}
\frac{m}{\bar{\rho}}n\left(m,z\right)dm=f\left(\nu\right)d\nu
,\end{equation}
where $\bar{\rho}$ is the comoving density of the background with
\begin{equation}
f\left(\nu\right)=A\left(p\right)\left(1+\left(q\nu\right)^{-p}\right)\sqrt{\frac{q}{2\pi\nu}}e^{-q\nu/2},\quad\nu=\frac{\delta_{SC}^{2}}{\sigma^{2}\left(m,z\right)}
.\end{equation}
In this formula, $A\left(p\right)\simeq0.322$ such that $\intop f\left(\nu\right)d\nu=1$,
$p\simeq0.3$, $q\simeq0.75$, $\delta_{SC}\simeq1.68$ is the critical
density required for spherical collapse, and
\begin{equation}
\sigma^{2}\left(m,z\right)=\int dk\frac{k^{2}P\left(k,z\right)}{2\pi^{2}}W\left(kR\right)^{2}
\end{equation}
is the variance of the initial density field extrapolated to the present
time using the linear prediction: $P\left(k,z\right)=\left(D^{+}\left(z\right)/D^{+}\left(0\right)\right)^{2}P\left(k,0\right)$. Here,
$D^{+}$ is the linear growth factor, $R=\left(3m/4\pi\bar{\rho}\right)^{1/3}$
is the radius of a sphere of mean comoving density $\bar{\rho}$ enclosing
a mass $m$, and $W$ is the Fourier transform of a top-hat function: $W\left(x\right)=3/x^{3}\left(\sin\left(x\right)-x\cos\left(x\right)\right)$.
To compute the values of the linear power spectrum, I use the CAMB
routines \citep{lewis2002cosmological}. The value $\nu=1$ defines
a characteristic mass scale $m_{*}\simeq4\times10^{12}M_{\text{\ensuremath{\odot}}}/h$
at $z=0$.

In practice, owing to the limited range of integration ($v\apprge10^{-2}$,
otherwise the integration would bring unphysical values of the mass),
the value of the parameter $A\left(p\right)$ is adapted to ensure
that $\int f\left(\nu\right)d\nu=1$.

\subsubsection{Halo biasing\label{sub:Halo-Biasing}}

Following \cite{mo1995ananalytic} (see also \cite{sheth1999largescale}
for an extension), the bias parameters of the overdensity of halos
relative to the matter overdensity $\delta_{h}\left(\boldsymbol{x,}z;m\right)=\sum_{k>0}b_{k}\left(m,z\right)\delta\left(\boldsymbol{x}\right)^{k}/k!$
in the spherical collapse model \citep{gunn1972onthe} are given by

\begin{eqnarray}
b_{1}(m,z) & = & 1+\varepsilon_{1}+E_{1},\nonumber \\
b_{2}(m,z) & = & 2(1+a_{2})(\varepsilon_{1}+E_{1})+\varepsilon_{2}+E_{2},\nonumber \\
b_{3}\left(m,z\right) & = & 6\left(a_{2}+a_{3}\right)\left(\varepsilon_{1}+E_{1}\right)+3\left(1+2a_{2}\right)\left(\varepsilon_{2}+E_{2}\right)+\varepsilon_{3}+E_{3}
\end{eqnarray}
with
\begin{eqnarray}
\varepsilon_{1} & = & \frac{q\nu-1}{\delta_{SC}},\quad\varepsilon_{2}=\frac{q\nu}{\delta_{SC}}\left(\frac{q\nu-3}{\delta_{SC}}\right),\quad\varepsilon_{3}=\frac{q\nu}{\delta_{SC}}\left(\frac{q\nu-3}{\delta_{SC}}\right)^{2}\nonumber \\
E_{2} & = & \frac{2p}{\delta_{SC}\left(1+\left(q\nu\right)^{p}\right)},\quad\frac{E_{2}}{E_{1}}=\frac{1+2p}{\delta_{SC}}+2\varepsilon_{1},\quad\frac{E_{3}}{E_{1}}=\frac{4\left(p^{2}-1\right)+6pq\nu}{\delta_{SC}^{2}}+3\varepsilon_{1}^{2}\nonumber \\
a_{2} & = & -17/21,\quad a_{3}=341/567
.\end{eqnarray}

These bias parameters obey the consistency relations 
\begin{equation}
\int dm\frac{m}{\bar{\rho}}n\left(m,z\right)b_{k}\left(m,z\right)=\delta_{k1}^{K}
.\end{equation}
Again, owing to the limited range of integration, the bias parameters
are rescaled so as to ensure the consistency relations.

Within this framework, it is easy to compute the halo-halo correlations
$P_{hh}$, $B_{hhh}$, and $T_{hhhh}$. The complete set of formulae for these
correlations are given in Appendix \ref{sec:powerspectrum}, with
corrections that complete the formulae presented in \cite{cooray2002halomodels}.

\subsubsection{Halo profiles\label{sub:Halo-Profiles}}

For the halo profile, I use a NFW profile \citep{navarro1997auniversal}
given by 
\begin{equation}
\rho\left(r;m\right)=\frac{\rho_{s}}{\left(r/r_{s}\right)\left(1+r/r_{s}\right)^{2}}
,\end{equation}
where $r_{s}=R_{vir}/c$, where $c$ is known as the concentration parameter,
and $R_{vir}$ is the virialization radius defined by $4\pi R_{vir}^{3}\Delta\bar{\rho}/3=m$
with $\Delta\left(z\right)=\left(18\pi^{2}+82\left(\Omega_{m}\left(z\right)-1\right)-39\left(\Omega_{m}\left(z\right)-1\right)^{2}\right)/\Omega_{m}\left(z\right)$
given by the spherical collapse model \citep{bryan1998statistical}.
The parameter $\rho_{s}$ is obtained by requesting that $m=\int d^{3}\boldsymbol{r}\rho\left(r;m\right)$.
For the concentration parameter I use \citep{bullock2001profiles}

\begin{equation}
c\left(m,z\right)=9\left(\frac{m}{m_{*}\left(z\right)}\right)^{-0.13}
.\end{equation}

In what follows I use the Fourier transform of the normalized
NFW profile $u\left(r;m\right)=\rho\left(r;m\right)/m$, which is
\begin{eqnarray}
u\left(k;m\right) & = & \frac{4\pi\rho_{s}r_{s}^{3}}{m}\left\{ \sin\left(kr_{s}\right)\left[\mathrm{Si}\left(\left[1+c\right]kr_{s}\right)-\mathrm{Si}\left(kr_{s}\right)\right]-\frac{\sin\left(ckr_{s}\right)}{\left(1+c\right)kr_{s}}\right.\nonumber \\
 & + & \left.\cos\left(kr_{s}\right)\left[\mathrm{Ci}\left(\left[1+c\right]kr_{s}\right)-\mathrm{Ci}\left(kr_{s}\right)\right]\vphantom{\frac{\sin\left(ckr_{s}\right)}{\left(1+c\right)kr_{s}}}\right\} 
,\end{eqnarray}
where the sine and cosine integrals are
\begin{equation}
\mathrm{Ci}\left(x\right)=-\intop_{x}^{+\infty}\frac{\cos t}{t}dt\quad and\quad\mathrm{Si}\left(x\right)=\intop_{0}^{x}\frac{\sin t}{t}dt
.\end{equation}

\subsubsection{The power spectrum and trispectrum in the halo model\label{sub:The-Power-Spectrum}}

In this section I follow the formalism developed by \cite{scherrer1991statistics}.
The comoving dark matter density field is written as 
\begin{eqnarray}
\rho_{tot}\left(\boldsymbol{x}\right) & = & \sum_{i}\rho\left(\boldsymbol{x}-\boldsymbol{x_{i}},m_{i}\right)\nonumber \\
 & \equiv & \sum_{i}m_{i}u\left(\boldsymbol{x}-\boldsymbol{x_{i}}\right)\nonumber \\
 & = & \sum_{i}\int dmd^{3}\boldsymbol{x'}\delta_{D}\left(m-m_{i}\right)\delta_{D}^{3}\left(\boldsymbol{x'}-\boldsymbol{x_{i}}\right)m\,u\left(\boldsymbol{x}-\boldsymbol{x'},\,m\right)
,\end{eqnarray}
where the sum is performed over the halos $i$ and $\rho$ is the
profile of a halo of mass $m_{i}$. The profile $u$ is defined such
that $\int d^{3}\boldsymbol{x}u\left(\boldsymbol{x}\right)=1$. The
number density of halos is
\begin{equation}
n\left(m\right)=\left\langle \sum_{i}\delta_{D}\left(m-m_{i}\right)\delta_{D}^{3}\left(\boldsymbol{x}-\boldsymbol{x_{i}}\right)\right\rangle 
,\end{equation}
such that
\begin{equation}
\bar{\rho}\equiv\left\langle \rho\left(\boldsymbol{x}\right)\right\rangle =\int mn\left(m\right)dm
.\end{equation}

It is then straightforward to compute the power spectrum in this model.
It can be split into two terms: the contributions coming from
the same halo and the ones coming from two different halos:
\begin{eqnarray}
P\left(k\right) & = & P^{1h}\left(k\right)+P^{2h}\left(k\right),\quad\mathrm{where}\nonumber \\
P^{1h}\left(k\right) & = & M_{02}\left(k,k\right)\nonumber \\
P^{2h}\left(k\right) & = & \left[M_{11}\left(k\right)\right]^{2}P^{PT}\left(k\right)
.\end{eqnarray}
 The power spectrum coming from two different halos is evaluated with
Equation (\ref{eq:Phh}). In this equation, the linear power spectrum
is used rather than the non-linear one. It is an approximation made
so as not to overestimate the power spectrum on intermediate and small
scales. On small scales, the one-halo term will dominate anyway. I
have introduced the convenient notation:
\begin{equation}
M_{ij}\left(k_{1},...,k_{j}\right)=\intop dmn\left(m\right)b_{i}\left(m\right)\left(\frac{m}{\bar{\rho}}\right)^{j}u\left(k_{1};m\right)u\left(k_{2};m\right)...u\left(k_{j};m\right)\label{eq:Mij}
.\end{equation}
These coefficients depend only on the norm of the $k_{i}$.
The trispectrum is evaluated in the same way. Since the entire formula
is a bit long, it is given in Appendix \ref{sec:powerspectrum}, with
corrections to the formula given in \cite{cooray2001powerspectrum}.

\section{Data\label{sec:Data}}

\subsection{Galaxy map}

In this study, I use the galaxies from the CFHTLenS\footnote{\href{http://cfhtlens.org}{http://cfhtlens.org}}
galaxy survey, a wide part of the Canada-France Hawaii Telescope Legacy
Survey (CFHTLS), which consists in four fields centred at $2^{\mathrm{h}}18^{\mathrm{m}}00^{\mathrm{s}}$
$-07^{\mathrm{d}}00^{\mathrm{m}}00^{\mathrm{s}}$ for W1, $08^{\mathrm{h}}54^{\mathrm{m}}00^{\mathrm{s}}$
$-04^{\mathrm{d}}15^{\mathrm{m}}00^{\mathrm{s}}$ for W2, $14^{\mathrm{h}}17^{\mathrm{m}}54^{\mathrm{s}}$
$+54^{\mathrm{d}}30^{\mathrm{m}}31^{\mathrm{s}}$ for W3, and $22^{\mathrm{h}}13^{\mathrm{m}}18^{\mathrm{s}}$
$+01^{\mathrm{d}}19^{\mathrm{m}}00^{\mathrm{s}}$ for W4. Each has
an area of $23-64\mathrm{\;deg^{2}}$, a total survey area of 154
$\mathrm{deg^{2}}$, and a depth in the $i_{AB}$ band of $i_{AB}=24.7$
\citep{heymans2012cfhtlens, erben2013cfhtlens}. The survey area was
imaged using the Megaprime wide field imager mounted at the prime
focus of the Canada-France-Hawaii Telescope (CFHT) and equipped with
the MegaCam camera. MegaCam comprises an array of $9\times4$ CCDs
and has a field of view of 1 $\mathrm{deg^{2}}$.

I limit my analysis to galaxies in the redshift range $0.2<z<1.3$.
 These galaxies have been confirmed to have a photometric redshift distribution
that resembles the measured spectroscopic redshift distribution \citep{heymans2012cfhtlens}.
Galaxies selected with $i_{AB}<24.5$ in this redshift slice have
a scatter of $0.03<\sigma_{\Delta z}<0.06$ (where $\sigma_{\Delta z}^{2}$
is the variance in the difference between the photometric and spectroscopic
redshifts $(z_{p}-z_{s})/(1+z_{s})$) with 10\% of the galaxies classified
as outliers \citep{benjamin2012cfhtlens}. The reduction pipeline
has been set to star\_flag = 0 and $\mathrm{mask\leq1}$ (description
of each flag can be found in \cite{erben2013cfhtlens}), and the magnitude
cut to $18.0<i_{AB}<24.0$. The resulting catalogue has a number of
galaxies $N_{gal}\simeq6.58\times10^{6}$ and a sky coverage of $f_{sky}\simeq3.5\times10^{-3}$.
Table \ref{tab:Catalogue} sums up the catalogue parameters in each
patch.

\begin{table}[h]
\begin{tabular}{|c||ccc|}
\hline 
Patch & $N_{gal}$ & $f_{sky}$ & $\bar{n}\,\left(\mathrm{gal\,arcmin^{-1}}\right)$\tabularnewline
\hline 
\hline 
All & $6.82\times10^{6}$ & $3.5\times10^{-3}$ & $15.1$\tabularnewline
W1 & $3.02\times10^{6}$ & $1.5\times10^{-3}$ & $15.0$\tabularnewline
W2 & $0.87\times10^{6}$ & $0.5\times10^{-3}$ & $14.8$\tabularnewline
W3 & $1.95\times10^{6}$ & $1.0\times10^{-3}$ & $15.3$\tabularnewline
W4 & $0.97\times10^{6}$ & $0.5\times10^{-3}$ & $15.3$\tabularnewline
\hline 
\end{tabular}

\protect\caption{CFHTLenS catalogue data\label{tab:Catalogue}}

\end{table}

\begin{figure}[h]
\includegraphics[scale=0.9]{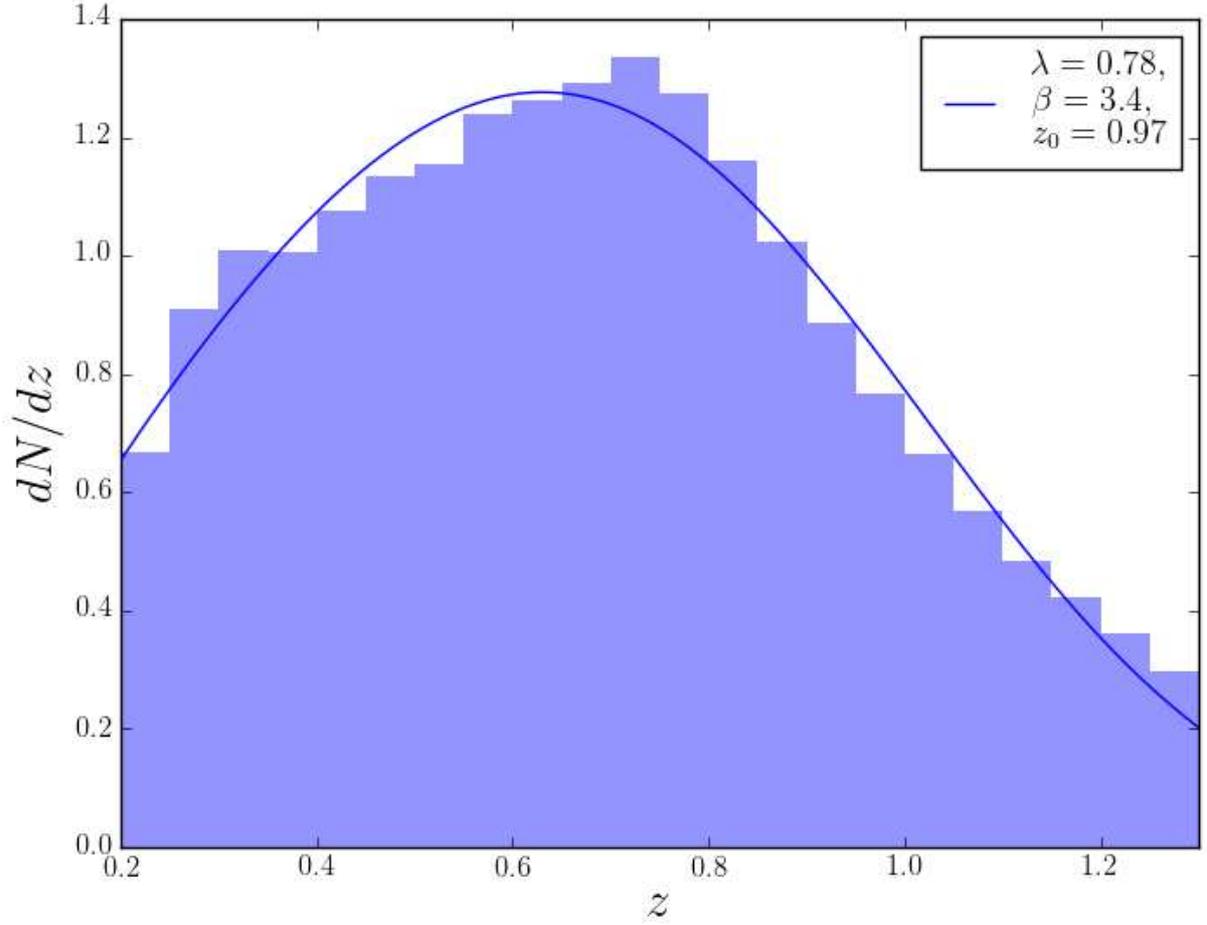}

\protect\caption{Redshift distribution of the CFHTLenS galaxies for all patches. The
histogram is the recovered redshift distribution obtained by averaging
the individual $P\left(z\right)$ and the solid line represents the
best-fit using Equation (\ref{eq:dNdz}). The best-fit values of $\lambda$,
$\beta$, and $z_{0}$ are used in Equation (\ref{eq:Wg}) for $W^{g}$.\label{fig:Redshift-distribution}}

\end{figure}

The redshift distribution $dN/dz$ used in Equation (\ref{eq:Wg})
is obtained by averaging the individual $P\left(z\right)$ of each
galaxy in each redshift bin, ranging from $0.2<z<1.3$ with a bin
size $\Delta z$ of $0.05$. This distribution is then fitted for
analytic convenience with an incomplete gamma distribution, omitting
the negligible error in redshift:
\begin{equation}
\frac{dN}{dz}=\frac{\alpha}{z_{0}}\left(\frac{z}{z_{0}}\right)^{\lambda}\exp\left(-\left(\frac{z}{z_{0}}\right)^{\beta}\right)\label{eq:dNdz}
\end{equation}
as shown in Figure \ref{fig:Redshift-distribution}. The best-fit
values for the galaxy sample are $\lambda=0.78$, $\beta=3.4$,
and $z_{0}=0.97$.

The galaxy overdensity is computed as a HEALPix\footnote{\href{http://healpix.jpl.nasa.gov}{http://healpix.jpl.nasa.gov}}
map \citep{gorski2005healpix} with $N_{side}=2048$ (corresponding
to a pixel size of $\sim1.'7$) by
\begin{equation}
\delta_{i}=\frac{N_{i}/w_{i}-\left\langle N\right\rangle }{\left\langle N\right\rangle }\label{eq:deltai}
,\end{equation}
where $w_{i}=S_{i}^{u}/S_{i}$ is the ratio between the unmasked surface
of a HEALPix pixel and the total suface of this pixel, $N_{i}$ is
the number of galaxies counted in a pixel, and $\left\langle N\right\rangle =\left(\sum_{i}N_{i}/w_{i}\right)/N_{pix}$
is the mean number of galaxies per pixel, corresponding to a number
density of $\bar{n}=15.1$ galaxies per square arcminute.

\subsection{Convergence map\label{sub:Convergence-map}}

I use the observed and simulated convergence map from the Planck 2015
release\footnote{\href{http://irsa.ipac.caltech.edu/data/Planck/release_2/ancillary-data/HFI_Products.html}{http://irsa.ipac.caltech.edu/data/Planck/release\_{}2/ancillary-data/HFI\_{}Products.html}}
\citep{planckcollaboration2015planck2015} and compare the results
with the 2013 release\footnote{\href{http://irsa.ipac.caltech.edu/data/Planck/release_1/ancillary-data/HFI_Products.html}{http://irsa.ipac.caltech.edu/data/Planck/release\_{}1/ancillary-data/HFI\_{}Products.html}}
\citep{planckcollaboration2014textitplanck}. The 2015 map is produced
by applying a quadratic estimator to all nine frequency bands with a
galaxy and point-source mask, leaving a total of 67.3\% of the sky
for analysis. The 2013 map is obtained by combining only the 143 and
217 GHz channels. The 2015 map is released as a $\kappa_{lm}$ map
for $8\leq l\leq2048$, while the 2013 map is a $\bar{\phi}$ map
and so is transformed into a convergence map by taking the transform
\begin{equation}
\kappa_{lm}=\frac{1}{R_{l}^{\phi\phi}}\frac{l\left(l+1\right)}{2}\bar{\phi}_{lm}
\end{equation}
where $R_{l}^{\phi\phi}$ is a normalization factor explained in \cite{planckcollaboration2014textitplanck}.
The 2013 and 2015 masks are combined for consistency. The final mask
for the cross-correlation is obtained by multiplying the convergence
mask and the galaxy mask, then applied by converting the convergence
multi-poles $\kappa_{lm}$ in real space and multiplying with the mask.

The noise for the auto-correlation has been estimated from the set
of 100 simulated lensing maps released by the Planck team. The noise
power spectrum $N_{l}^{\kappa\kappa}$ was estimated by averaging
over the 100 simulations the autospectrum of the masked difference
map between the reconstructed and the input lensing map.

\section{Constraints on galaxy bias and cross-correlation amplitude\label{sec:Constraints-on-Galaxy}}

Following \cite{planckcollaboration2014textitplanck} and \cite{planckcollaboration2015planck2015},
I introduce a new parameter, $A$, named the lensing amplitude, that
scales the amplitude of the cross-correlation: $\tilde{C}_{l_{i}}^{\kappa g}=AC_{l_{i}}^{\kappa g}\left(b\right)$.
Its expected value is obviously one. I then use the combined cross-
and auto-correlation to constrain the galaxy bias $b$ and the lensing
amplitude $A$.

I use the following scheme, developed by e.g. \cite{bianchini2015crosscorrelation} : the cross and auto-correlation are organized
in a single vector following
\begin{equation}
\boldsymbol{C}=\left(\boldsymbol{C}^{\kappa g},\boldsymbol{C}^{gg}\right)
\end{equation}
where $\boldsymbol{C}^{XY}$ is the vector containing $\left(C_{l_{i}}^{XY}\right)^{i=1..N_{bin}}$
in the $N_{bin}=19$ bins used. The total covariance matrix writes
as
\begin{equation}
\Sigma=\left(\begin{array}{cc}
\Sigma^{\kappa g} & \left(\Sigma^{\kappa g-gg}\right)^{T}\\
\Sigma^{\kappa g-gg} & \Sigma^{gg}
\end{array}\right)
,\end{equation}
where the mixed covariance $\Sigma_{ij}^{\kappa g-gg}$ is defined
in Equation (\ref{eq:sigmakg-gg}). The covariances are evaluated
for the fiducial values $A=1$ and $b=1$. According to Bayes' theorem,
the posterior probability of a given set of parameters $\left(A,b\right)$
given the data $\boldsymbol{C}$, is 
\begin{equation}
P\left(A,b\mid\boldsymbol{C}\right)=\frac{P\left(A,b\right)P\left(\boldsymbol{C}\mid A,b\right)}{P\left(\boldsymbol{C}\right)}
\end{equation}
 where $P\left(A,b\right)$ is the prior on the parameters, $P\left(\boldsymbol{C}\mid A,b\right)$
is the likelihood function for measuring $\boldsymbol{C}$ given $A,b$,
and $P\left(\boldsymbol{C}\right)$ is a normalization factor. I assume
a Gaussian likelihood function, which takes the form 
\begin{equation}
P\left(\boldsymbol{C}\mid A,b\right)=\frac{1}{\sqrt{\left(2\pi\right)^{2N_{bin}}\det\left(\Sigma\right)}}\exp\left\{ -\frac{1}{2}\left(\tilde{\boldsymbol{C}}-\boldsymbol{C}\left(A,b\right)\right)\Sigma^{-1}\left(\tilde{\boldsymbol{C}}-\boldsymbol{C}\left(A,b\right)\right)\right\} 
,\end{equation}
and a flat prior. To sample the parameter space, I use a
Monte-Carlo Markov Chain (MCMC) method employing the Python module
EMCEE \citep{foreman-mackey2013emceethe}, which is a public implementation
of the affine invariant MCMC ensemble sampler \citep{goodman2010ensemble}.
The resulting parameters are estimated by the median of the posterior
distribution after marginalizing over the other parameters with uncertainties
given by the $16^{th}$ and $84^{th}$ percentiles. For a Gaussian
distribution, the median is equal to the maximum likelihood value, and
the percentiles correspond to the $-1\sigma$ and $+1\sigma$ error
bars; here, the recovered distributions are very close to Gaussians.

\begin{figure}[h]
\includegraphics[width=0.6\paperwidth]{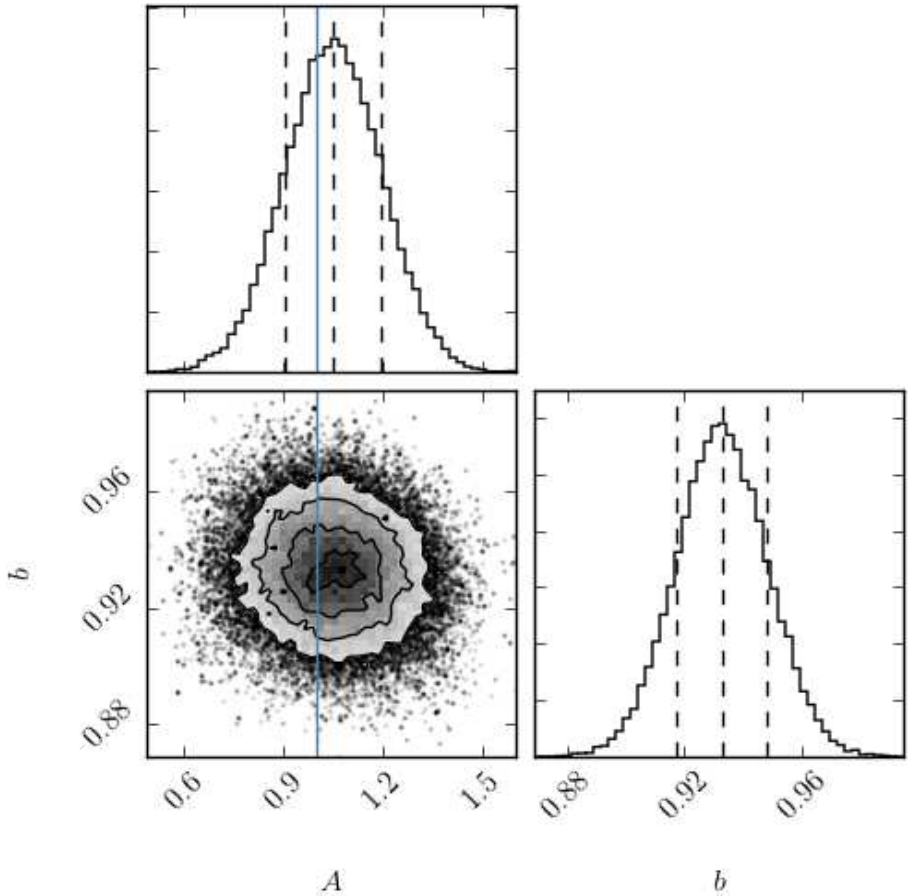}\protect\caption{Posterior distribution in the $A-b$ plane, together with the marginalized
distributions for each parameter, for the 2013 data. The contours
show the $0.5\sigma$, $1\sigma$, $1.5\sigma$, and $2\sigma$ lines
from the centre to the border. The solid red line represents the fiducial
value $A=1$, and the dashed lines the $-1\sigma$ and $+1\sigma$
error bars. \label{fig:Distrib_2013}}

\end{figure}

\begin{figure}[h]
\includegraphics[width=0.6\paperwidth]{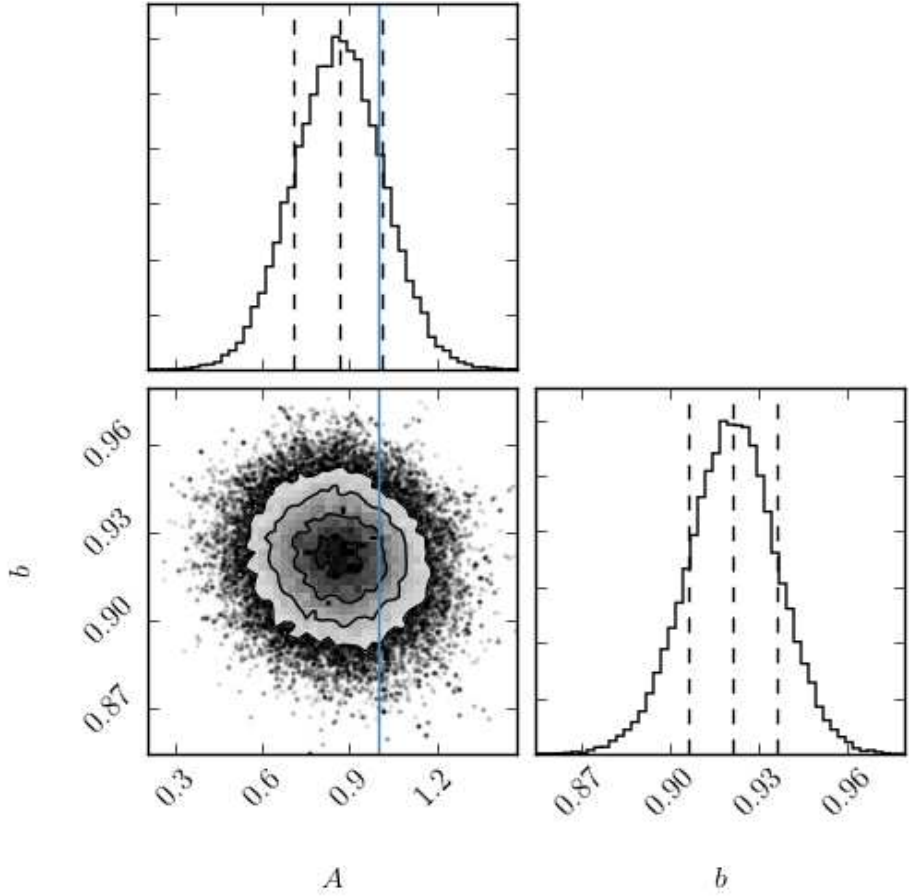}\protect\caption{Same plot as in Figure \ref{fig:Distrib_2013}, but for the 2015 data.\label{fig:Distrib_2015}}

\end{figure}

The two-dimensional posterior distribution and the marginalized ones
are shown in Figure \ref{fig:Distrib_2013} for the 2013 release,
and in Figure \ref{fig:Distrib_2015} for the 2015 release, using
all the patches together. Table \ref{tab:fit} sums up the values
of parameters $A$ and $b$ for each patch and for the two releases
with $1\sigma$ error bars, together with the $\chi^{2}$ calculated
as $\chi^{2}=\left(\tilde{\boldsymbol{C}}-\boldsymbol{C}\left(A,b\right)\right)\Sigma^{-1}\left(\tilde{\boldsymbol{C}}-\boldsymbol{C}\left(A,b\right)\right)$
for $\nu=2N_{bin}-2$ degrees of freedom. Parameter $b$ is mostly
constrained by the galaxy auto-correlation, so it has approximately
the same value for the two releases.

\begin{table}[h]
\begin{tabular}{|c|ccc||ccc|}
\hline 
 &  & 2013 &  &  & 2015 & \tabularnewline
Patch & $A$ & $b$ & $\chi^{2}/\nu$ & $A$ & $b$ & $\chi^{2}/\nu$\tabularnewline
\hline 
\hline 
All & $1.05_{-0.15}^{+0.15}$ & $0.93_{-0.02}^{+0.02}$ & $47.2/36$ & $0.86_{-0.16}^{+0.15}$ & $0.92_{-0.02}^{+0.02}$ & $37.4/36$\tabularnewline
\hline 
W1 & $0.69_{-0.22}^{+0.23}$ & $0.90_{-0.02}^{+0.02}$ & $54.3/36$ & $0.41_{-0.24}^{+0.24}$ & $0.89_{-0.03}^{+0.02}$ & $48.2/36$\tabularnewline
W2 & $1.29_{-0.34}^{+0.34}$ & $1.04_{-0.04}^{+0.04}$ & $29.0/36$ & $1.22_{-0.36}^{+0.35}$ & $1.03_{-0.04}^{+0.04}$ & $30.9/36$\tabularnewline
W3 & $1.34_{-0.28}^{+0.28}$ & $0.94_{-0.03}^{+0.03}$ & $50.9/36$ & $0.99_{-0.29}^{+0.29}$ & $0.92_{-0.03}^{+0.03}$ & $39.6/36$\tabularnewline
W4 & $1.28_{-0.40}^{+0.40}$ & $0.89_{-0.04}^{+0.04}$ & $44.1/36$ & $1.60_{-0.40}^{+0.42}$ & $0.88_{-0.04}^{+0.04}$ & $59.1/36$\tabularnewline
\hline 
\end{tabular}

\protect\caption{Best-fit values for $A$ and $b$ using both cross- and auto-correlation
for the 2013 and 2015 releases.\label{tab:fit}}
\end{table}

\begin{figure}[h]
\includegraphics[width=0.8\paperwidth]{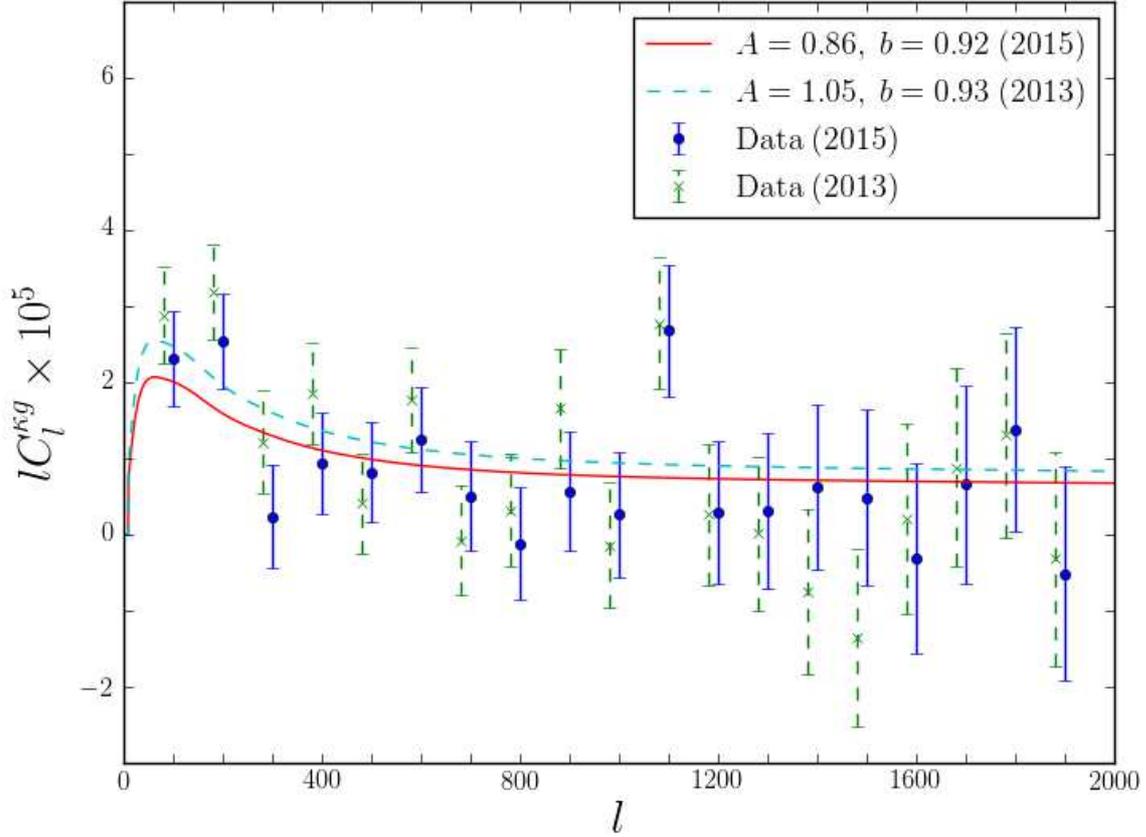}\protect\caption{CMB convergence - galaxy over-density cross-correlation $C_{l}^{\kappa g}$
using all patches. Error bars are computed from the theoretical correlation
matrix of Equation (\ref{eq:variance}). The solid line and
the solid blue points are the 2015 best-fit and the 2015
data, respectively, while the dashed line and the dashed green points
are the 2013 best-fit and the 2013 data.\label{fig:Clkg}}
\end{figure}

\begin{figure}[h]
\includegraphics[width=0.8\paperwidth]{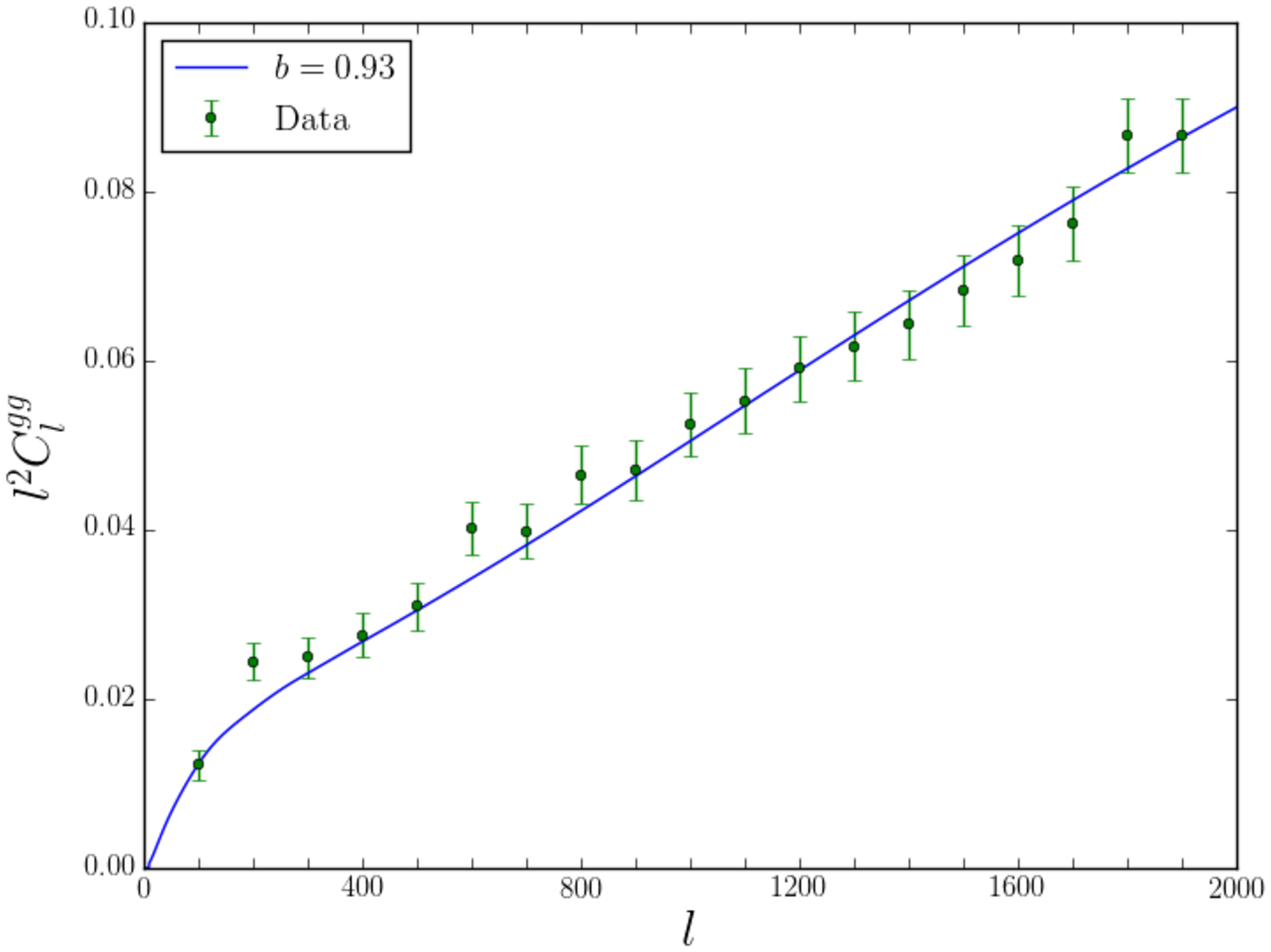}\protect\caption{As in Figure \ref{fig:Clkg}, but for the galaxy auto-correlation
$C_{l}^{gg}$ using all patches. The solid line is the best fit of
2013 and 2015 (which give the same value for $b$), and the data are
shown in green. \label{fig:Clgg}}

\end{figure}

The cross- and auto-correlations, together with the best-fit parameters
theoretical prediction (with all patches together), are shown in Figures
\ref{fig:Clkg} and \ref{fig:Clgg}. 

\section{Consistency checks\label{sec:Consistency-Checks}}

\subsection{Simulations\label{sec:Simulations}}

To validate the algorithm employed to reconstruct the convergence
and galaxy maps and to check the consistency of the theoretical error
bars in the Gaussian limit, I created $N_{sim}=100$ simulations of
the galaxy and convergence maps. I used the Healpy synfast function to
generate a set of $\kappa_{lm}$ and $g_{lm}$ and the theoretical
$C_{l}^{\kappa\kappa}$, $C_{l}^{\kappa g}$, and $C_{l}^{gg}$ given
in Equations (\ref{eq:Clkk}), (\ref{eq:Clkg}), and (\ref{eq:Clgg}),
with $b=1$. The multi-pole coefficients are synthetized by \begin{eqnarray}
\kappa_{lm} & = & \xi_{1}\left(C_{l}^{\kappa\kappa}\right)^{1/2}\nonumber \\
g_{lm} & = & \xi_{1}\frac{C_{l}^{\kappa g}}{\left(C_{l}^{\kappa\kappa}\right)^{1/2}}+\xi_{2}\left(C_{l}^{gg}-\frac{\left(C_{l}^{\kappa g}\right)^{2}}{C_{l}^{\kappa\kappa}}\right)^{2}
,\end{eqnarray}
where for each value of $l$ and $m>0$, $\xi_{1}$ and $\xi_{2}$
are two complex numbers drawn from a Gaussian distribution of mean
$0$ and variance $1$, whereas for $m=0$ they are real.

To account for the level of noise in the maps, I replaced
$C_{l}^{\kappa\kappa}$ by $C_{l}^{\kappa\kappa}+N_{l}^{\kappa\kappa}$
with $N_{l}^{\kappa\kappa}$ the noise given in the Planck Collaboration
release. As pointed out by the Planck 2013 Wiki\footnote{\href{http://wiki.cosmos.esa.int/planckpla/index.php/Main_Page}{http://wiki.cosmos.esa.int/planckpla/index.php/Main\_{}Page}},
this noise is not accurate enough for the auto-correlation, but it
should be sufficient for the cross-correlation, which is not biased
by this noise term. The noise in the galaxy map was accounted for by drawing
the number of galaxies in each pixel from a Poisson distribution with
mean
\begin{equation}
\lambda\left(\boldsymbol{\theta}\right)=\left\langle N\right\rangle \left(1+\delta\left(\boldsymbol{\theta}\right)\right)
\end{equation}
where $\left\langle N\right\rangle $ is the mean number of galaxies
per pixel of the original map, and $\delta\left(\boldsymbol{\theta}\right)$
 the simulated overdensity map. I then replaced the galaxy number
count $N_{i}/w_{i}$ in Equation \ref{eq:deltai} by $\lambda\left(\boldsymbol{\theta}\right)$.
The Poisson noise of variance $N_{l}^{gg}=1/\bar{n}$ was thus included
in this map with $\bar{n}$ the number of galaxies per steradian.

\begin{figure}[h]
\includegraphics[width=0.8\paperwidth]{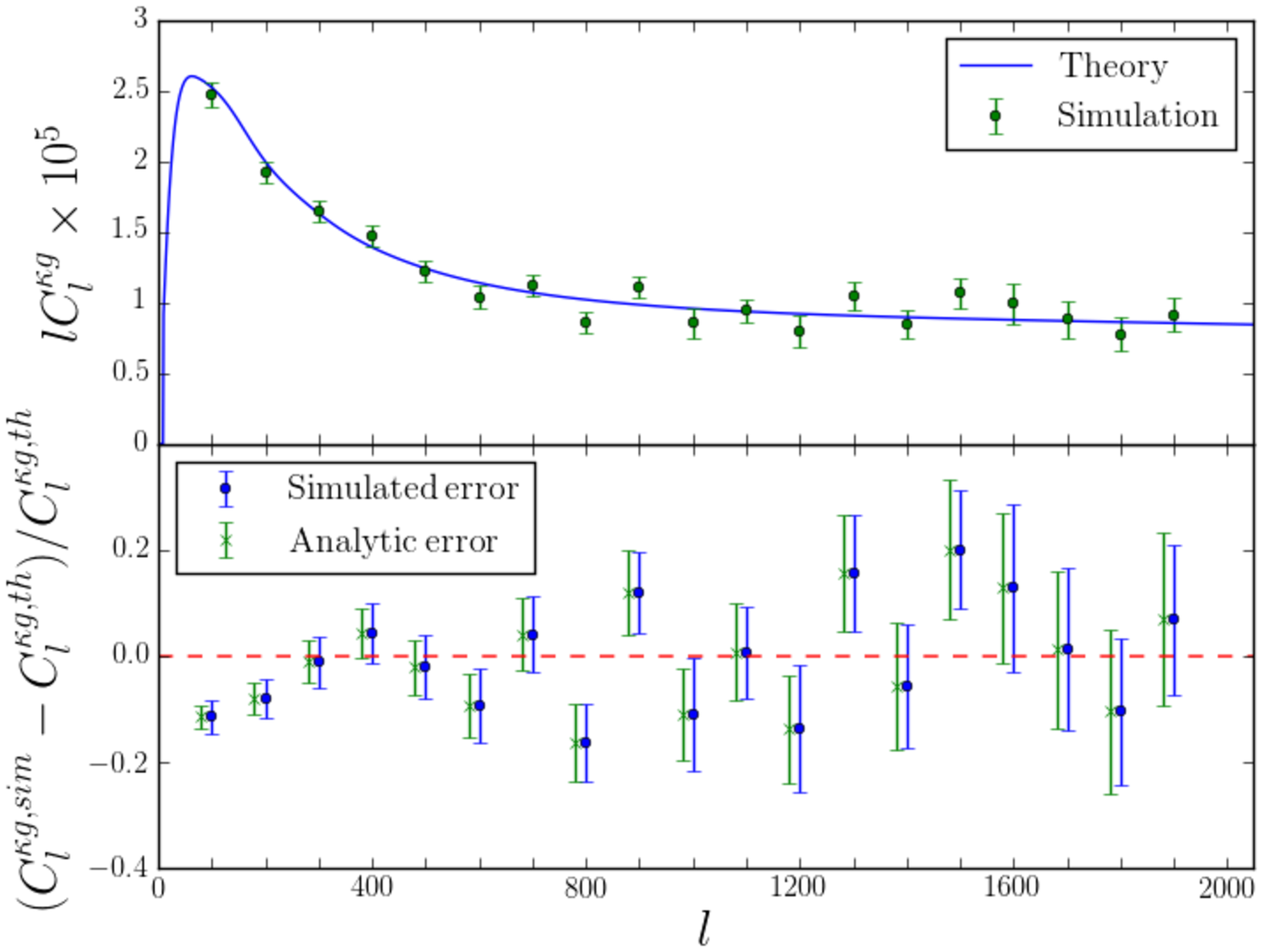}

\protect\caption{\emph{Upper panel}: Cross-correlation of the simulated galaxy and
lensing maps using $b=1$. The solid line represents the input cross-correlation,
and the points represent the reconstructed cross-correlation averaged
over $N_{sim}=100$ simulations, together with the simulated error
bars.\protect \\
\emph{Lower panel}: Fractional difference between the input and the
recovered cross-correlations. The blue error bars are recovered from
the simulations using Equation (\ref{eq:variance_simu}) for the covariance
matrix, and the red ones are analytic using Equation (\ref{eq:variance})
for the covariance, keeping only the Gaussian term.\label{fig:Clkg_simu}}
\end{figure}

\begin{figure}[h]
\includegraphics[width=0.8\paperwidth]{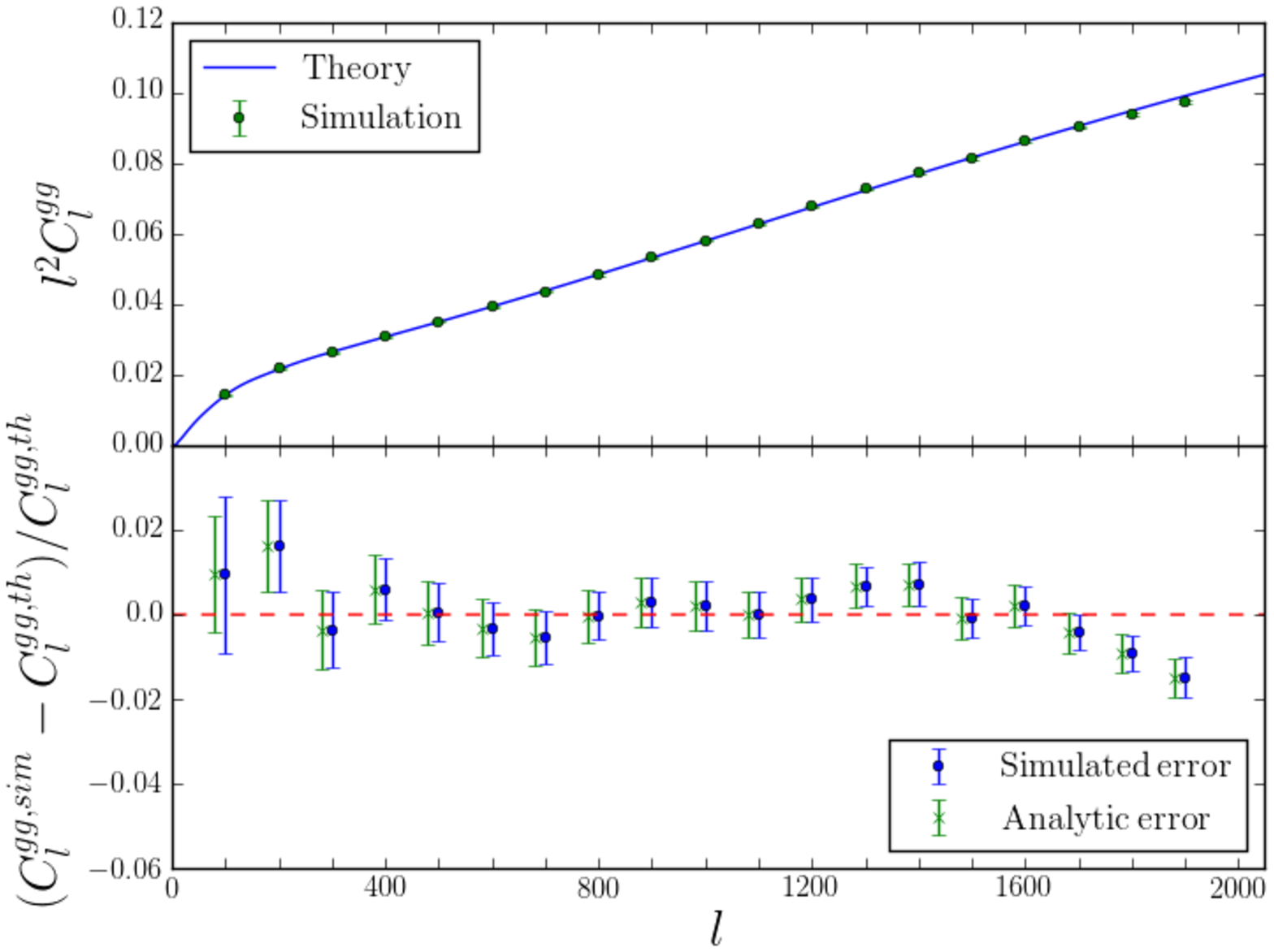}\protect\caption{Same plot as Figure (\ref{fig:Clkg_simu}), but for $C_{l}^{gg}$.\label{fig:Clgg_simu}}
\end{figure}

I then applied the spectral estimators described in the previous sections
to the simulated convergence and over-density maps. The recovered $C_{l}^{\kappa g}$
and $C_{l}^{gg}$ averaged over the 100 simulations are shown in Figs.
\ref{fig:Clkg_simu} and \ref{fig:Clgg_simu}. The mean correlation
was computed as 
\begin{equation}
\left\langle \tilde{C}_{l_{i}}^{XY}\right\rangle =\frac{1}{N_{sim}}\sum_{\alpha=1}^{N_{sim}}\tilde{C}_{l_{i}}^{XY,\alpha}
\end{equation}
where $X,Y=\left\{ \kappa,g\right\} $, $\alpha$ is the number of
the simulation, and $i$ refers to the bin in $l$ of width $\delta l=100$.
The covariance matrix of the samples was computed as
\begin{equation}
\tilde{\Sigma}_{ij}^{XY}=\frac{1}{N_{sim}-1}\sum_{\alpha=1}^{N_{sim}}\left(\tilde{C}_{l_{i}}^{XY,\alpha}-\left\langle \tilde{C}_{l_{i}}^{XY}\right\rangle \right)\left(\tilde{C}_{l_{j}}^{XY,\alpha}-\left\langle \tilde{C}_{l_{j}}^{XY}\right\rangle \right)\label{eq:variance_simu}
.\end{equation}

For comparison, I show the theoretical error bars and the recovered
simulated error bars for the mean correlation computed as 
\begin{equation}
\Delta\tilde{C}_{l_{i}}^{XY}=\left(\frac{\tilde{\Sigma}_{ii}^{XY}}{N_{sim}}\right)^{1/2}
\end{equation}
with the same formula for the theoretical error bars, using only the
Gaussian term in Equation (\ref{eq:variance}) because the simulations
are Gaussian. They are in very good agreement.

\begin{figure}[h]
\includegraphics[width=0.6\paperwidth]{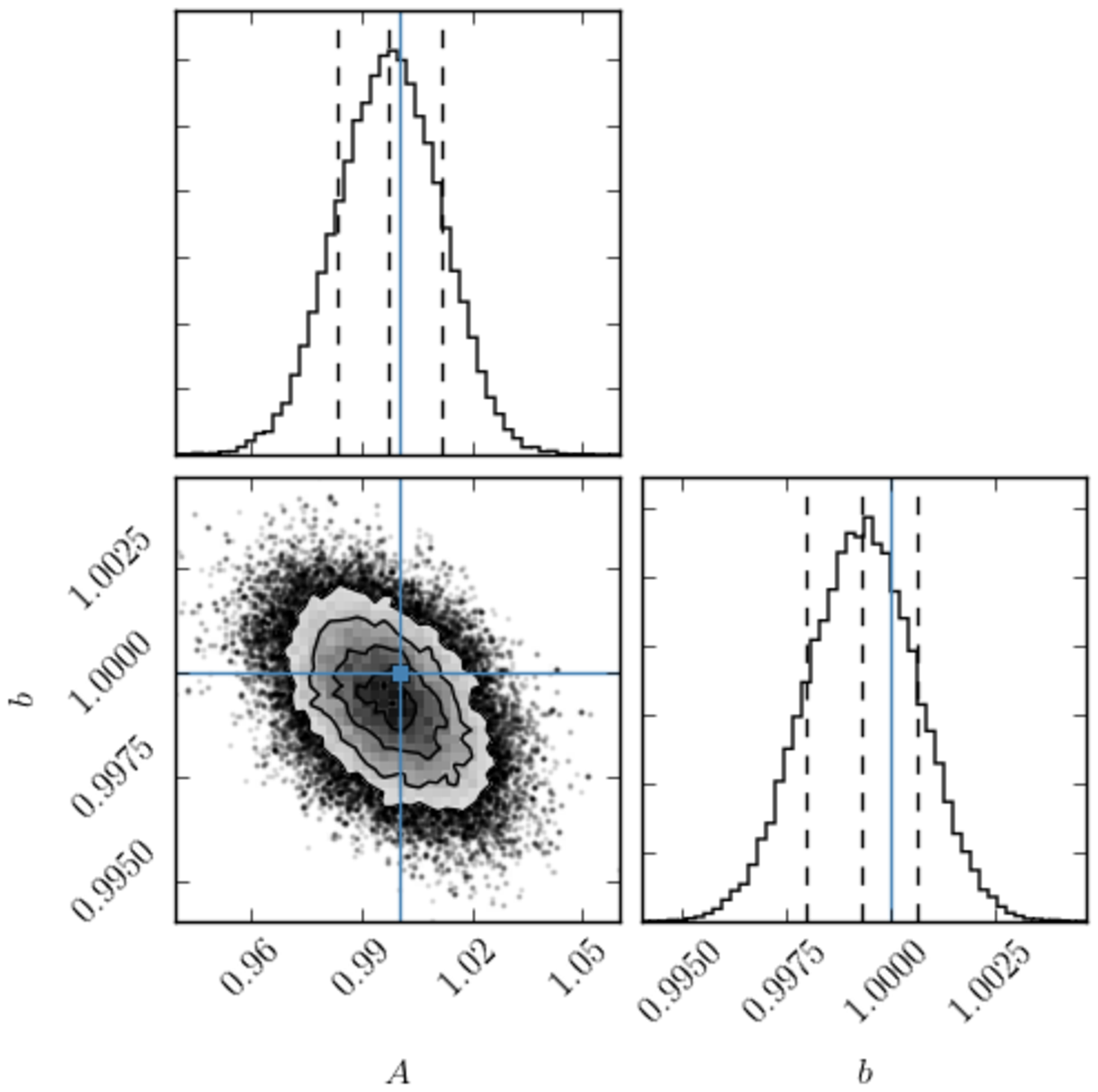}

\protect\caption{As in Figure \ref{fig:Distrib_2013}, but for the Gaussian simulations.
The horizontal and vertical solid lines show the fiducial values $A=1$
and $b=1$ used in the simulations.\label{fig:Distrib-Simu}}

\end{figure}

I also fitted a galaxy bias and a lensing amplitude following the pipeline
explained in Section \ref{sec:Constraints-on-Galaxy} and compared
them to the fiducial values used in the simulations $A=1$ and $b=1$.
To this aim I replaced the correlations $C_{l_{i}}^{XY}$ by the mean
correlations $\left\langle \tilde{C}_{l_{i}}^{XY}\right\rangle $
and the covariances matrices $\Sigma^{XY}$ by the mean correlation
covariance $\tilde{\Sigma}^{XY}/N_{sim}$. The fitted values of the
parameters are $A=0.998_{-0.014}^{+0.014}$ and $b=0.999_{-0.001}^{+0.001}$,
indicating that the reconstruction is good. Figure \ref{fig:Distrib-Simu}
shows a corner plot of the MCMC sampler for the simulations, as in
Figure \ref{fig:Distrib_2013}.

\subsection{Null tests\label{sub:Null-Tests}}

\begin{figure}[h]
\includegraphics[width=0.8\paperwidth]{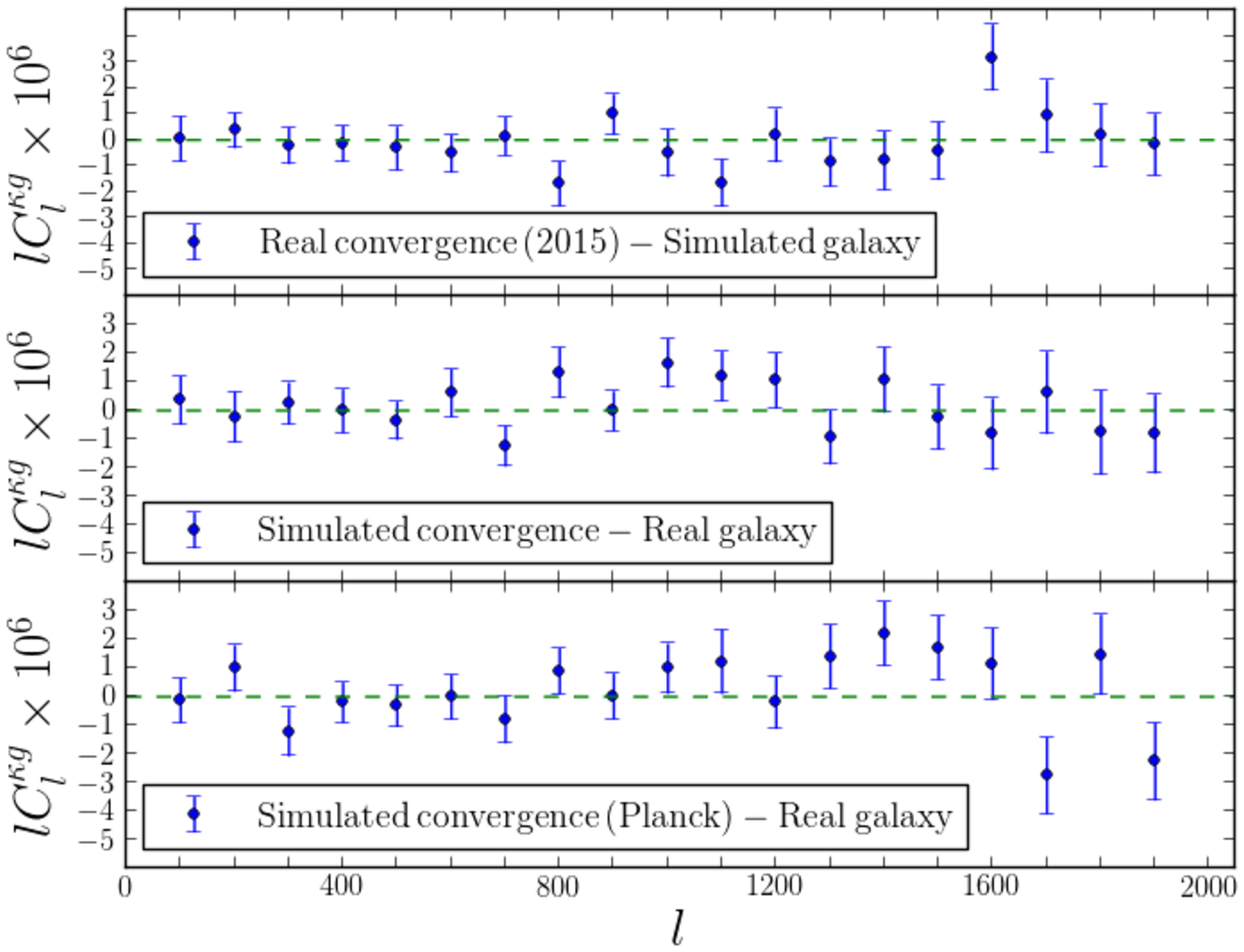}\protect\caption{\emph{Upper panel}: Mean correlation between the true lensing map
of 2015 and 100 simulated galaxy maps using all patches.\protect \\
\emph{Middle panel}: Mean correlation between the true galaxy map
using all patches and 100 gaussian simulated lensing maps.\protect \\
\emph{Lower panel: }Mean correlation between the true galaxy map
using all patches and 100 Planck simulated lensing maps.\protect \\
In all cases the signal is consistent with no correlation. \textbf{\label{fig:NullTest}}}
\end{figure}

To check that there is no systematics in the pipeline descibed above,
I performed a series of null tests consisting in cross-correlating
a real map with the $N_{sim}=100$ simulated maps of Section \ref{sec:Simulations},
both for a real convergence map with a simulated galaxy map and for
a simulated convergence map with a real galaxy map, using all the
patches together. The expected signal is of null amplitude, with a
simulated covariance given by Equation (\ref{eq:variance_simu}) applied
to the null test simulations. As shown in Figure \ref{fig:NullTest},
in both cases no significant signal was detected. The fitted values
of the product $A\times b$ (which is the amplitude of the cross-correlation)
are summarized in Table \ref{tab:NullTest}.

To validate the use of Gaussian simulations in the computation of
covariances, I also cross-correlated the 100 simulated Planck maps
given by the Planck collaboration with the real galaxy overdensity
field. I obtain a result similar to the Gaussian simulation null test
for the error bars, indicating that the use of Gaussian simulations
is relevant.

\begin{table}[h]
\begin{tabular}{|c|cc|}
\hline 
Null test & $A\times b$ & $\chi^{2}/\nu$\tabularnewline
\hline 
\hline 
\multirow{1}{*}{2013 Real convergence - Simulated galaxy} & $-0.013_{-0.018}^{+0.018}$ & $24.8/19$\tabularnewline
\multirow{1}{*}{2015 Real convergence - Simulated Galaxy} & $-0.001_{-0.018}^{+0.017}$ & $21.0/19$\tabularnewline
Simulated convergence - Real galaxy & $0.015_{-0.016}^{+0.016}$ & $21.5/19$\tabularnewline
Simulated convergence (Planck) - Real galaxy & $0.003_{-0.014}^{+0.014}$ & $25.3/19$\tabularnewline
\hline 
\end{tabular}

\protect\caption{Null tests performed as explained in Section \ref{sub:Null-Tests}.\label{tab:NullTest}}
\end{table}

\section{Summary and conclusions\label{sec:Summary-and-Conclusions}}

I have presented the results from a joint analysis of the cross and
auto-correlation of the CFHTLenS galaxy catalogue and Planck CMB lensing,
checking the reliability of my result by null tests and
Gaussian simulations. I found a galaxy bias of $b=0.92_{-0.02}^{+0.02}$
and a cross-correlation amplitude of $A=0.86_{-0.15}^{+0.15}$ for the
2015 release, whereas for the 2013 release I found $b=0.93_{-0.02}^{+0.02}$
and $A=1.05_{-0.15}^{+0.15}$. This confirms the difference between
the two releases shown by \cite{omori2015crosscorrelation}, but the
trend is less clear, and both results are compatible with each other.
These results are consistent with the amplitudes obtained from the
Planck lensing autocorrelation in \cite{planckcollaboration2015planck2015}: $A=0.987\pm0.025$ for the 2015 release and $A=1.005\pm0.043$ for
the 2013 release. The value of the galaxy bias suggests that galaxies
in this magnitude range are unbiased tracers of dark matter.

This study suggests that the joint analysis of the cross and auto-correlation
can put strong constraints on the properties of tracers of dark matter.
Forthcoming wide galaxy surveys probing fainter magnitudes will improve
the constraining power of this kind of study, both by a larger survey
area to improve statistics and a sufficient galaxy number density
to avoid shot noise. They are expected to put better constraints on
the cosmological model used in this paper.
\begin{acknowledgements}
I would first like to thank Simon Prunet for all the invaluable knowledge
he shared with me. This work is based on observations with
MegaPrime/MegaCam, a joint project of CFHT and CEA/DAPNIA, at the
Canada-France-Hawaii Telescope (CFHT), which is operated by the National
Research Council (NRC) of Canada, the Institut National des Sciences
de l'Univers of the Centre National de la Recherche Scientifique (CNRS)
of France, and the University of Hawaii. This research used the facilities
of the Canadian Astronomy Data Centre operated by the National Research
Council of Canada with the support of the Canadian Space Agency. CFHTLenS
data processing was made possible thanks to significant computing
support from the NSERC Research Tools and Instruments grant programme.
Some of the results in this paper have been derived using the HEALPix
\citep{gorski2005healpix} package.
\end{acknowledgements}

\begin{appendix}

\section{The halo model power spectrum, bispectrum, and trispectrum\label{sec:powerspectrum}}

Using the bias parameters given in Section \ref{sub:Halo-Biasing},
the halo-halo correlations are (the dependence on $z$ is omitted
for clarity, along with the $k$'s and $m$'s when not needed):

\begin{eqnarray}
P_{hh} & = & b_{1}\left(m_{1}\right)b_{1}\left(m_{2}\right)P^{PT}\left(k\right),\nonumber \\
B_{hhh} & = & b_{1}\left(m_{1}\right)b_{1}\left(m_{2}\right)b_{1}\left(m_{3}\right)B^{PT}+\left[b_{2}\left(m_{1}\right)b_{1}\left(m_{2}\right)b_{1}\left(m_{3}\right)P^{PT}\left(k_{2}\right)P^{PT}\left(k_{3}\right)+cyc.\right],\nonumber \\
T_{hhhh} & = & b_{1}\left(m_{1}\right)b_{1}\left(m_{2}\right)b_{1}\left(m_{3}\right)b_{1}\left(m_{4}\right)T^{PT}\nonumber \\
 & + & b_{3}\left(m_{1}\right)b_{1}\left(m_{2}\right)b_{1}\left(m_{3}\right)b_{1}\left(m_{4}\right)P^{PT}\left(k_{2}\right)P^{PT}\left(k_{3}\right)P^{PT}\left(k_{4}\right)+cyc.\nonumber \\
 & + & b_{2}\left(m_{1}\right)b_{1}\left(m_{2}\right)b_{1}\left(m_{3}\right)b_{1}\left(m_{4}\right)\left[P^{PT}\left(k_{2}\right)B^{PT}\left(\boldsymbol{k_{1}}+\boldsymbol{k_{2}},\boldsymbol{k_{3}},\boldsymbol{k_{4}}\right)+\left(2\leftrightarrow3\right)+\left(2\leftrightarrow4\right)\right]+cyc.\nonumber \\
 & + & b_{2}\left(m_{1}\right)b_{1}\left(m_{3}\right)b_{1}\left(m_{4}\right)\left\{ b_{2}\left(m_{2}\right)P^{PT}\left(k_{3}\right)P^{PT}\left(k_{4}\right)\left[P^{PT}\left(\left|\boldsymbol{k_{1}}+\boldsymbol{k_{3}}\right|\right)+P^{PT}\left(\left|\boldsymbol{k_{1}}+\boldsymbol{k_{4}}\right|\right)\right]\right.\nonumber \\
 &  & \hphantom{b_{2}\left(m_{1}\right)b_{1}\left(m_{3}\right)b_{1}\left(m_{4}\right)}\left.+\left(2\leftrightarrow3\right)+\left(2\leftrightarrow4\right)\vphantom{P^{PT}\left(\left|\boldsymbol{k_{1}}+\boldsymbol{k_{3}}\right|\right)}\right\} +cyc\label{eq:Phh}
.\end{eqnarray}
The formula given in \cite{cooray2002halomodels} (Equation
(90)) for the trispectrum is actually incomplete. Here, $P^{PT}$, $B^{PT}$
and $T^{PT}$ are the power spectrum, bispectrum, and trispectrum at
lowest order in perturbation theory (see \cite{bernardeau2002largescale}
for a review) given by \begin{eqnarray}
B^{PT}\left(\boldsymbol{k_{1}},\boldsymbol{k_{2}},\boldsymbol{k_{3}}\right) & = & 2F_{2}^{s}\left(\boldsymbol{k_{1}},\boldsymbol{k_{2}}\right)P^{PT}\left(k_{1}\right)P^{PT}\left(k_{2}\right)+\left(\boldsymbol{k_{1}}\leftrightarrow\boldsymbol{k_{3}}\right)+\left(\boldsymbol{k_{2}}\leftrightarrow\boldsymbol{k_{3}}\right)\nonumber \\
T^{PT}\left(\boldsymbol{k_{1}},-\boldsymbol{k_{1}},\boldsymbol{k_{2}},-\boldsymbol{k_{2}}\right) & = & 4P^{PT}\left(\boldsymbol{k_{1}}+\boldsymbol{k_{2}}\right)\left[F_{2}^{s}\left(-\boldsymbol{k_{1}},\boldsymbol{k_{1}}+\boldsymbol{k_{2}}\right)P^{PT}\left(k_{1}\right)+\left(\boldsymbol{k_{1}}\leftrightarrow\boldsymbol{k_{2}}\right)\right]+\left(\boldsymbol{k_{1}}\leftrightarrow-\boldsymbol{k_{1}}\right)\nonumber \\
 & + & 12\left[F_{3}^{s}\left(\boldsymbol{k_{1}},-\boldsymbol{k_{1}},\boldsymbol{k_{2}}\right)P^{PT}\left(k_{1}\right)^{2}P^{PT}\left(k_{2}\right)+\left(\boldsymbol{k_{1}}\leftrightarrow\boldsymbol{k_{2}}\right)\right]\label{eq:PT}
\end{eqnarray}
where the symmetrized kernels $F_{n}^{s}$ are derived in \cite{goroff1986coupling}.
There is a very small dependence of the kernels on the parameter $\Omega_{m}$,
which is ignored in this study.

The halo model power spectrum and trispectrum are then computed as
(using the notation $M_{ij}\left(k_{1},...,k_{j}\right)$ given in
Equation (\ref{eq:Mij})):
\begin{eqnarray}
P\left(k\right) & = & P^{1h}\left(k\right)+P^{2h}\left(k\right),\quad\mathrm{where}\nonumber \\
P^{1h}\left(k\right) & = & M_{02}\left(k,k\right)\nonumber \\
P^{2h}\left(k\right) & = & \left[M_{11}\left(k\right)\right]^{2}P^{PT}\left(k\right)
.\end{eqnarray}

Since I only need terms of the form $T\left(\boldsymbol{k_{1}},-\boldsymbol{k_{1}},\boldsymbol{k_{2}},-\boldsymbol{k_{2}}\right)$
(see Equation (\ref{eq:Tbarij})), the trispectrum can be simplified
as 
\begin{eqnarray}
T_{hhhh}\left(\boldsymbol{k_{1}},-\boldsymbol{k_{1}},\boldsymbol{k_{2}},-\boldsymbol{k_{2}}\right) & = & T^{1h}+T^{2h}+T^{3h}+T^{4h}
\end{eqnarray}
with
\begin{eqnarray}
T^{1h} & = & M_{04}\left(k_{1},k_{1},k_{2},k_{2}\right)\nonumber \\
T^{2h} & = & 2M_{11}\left(k_{1}\right)M_{13}\left(k_{1},k_{2},k_{2}\right)P^{PT}\left(k_{1}\right)+\left(k_{1}\leftrightarrow k_{2}\right)\nonumber \\
 & + & M_{12}\left(k_{1},k_{2}\right)^{2}\left[P^{PT}\left(\left|\boldsymbol{k_{1}}+\boldsymbol{k_{2}}\right|\right)+\left(\boldsymbol{k_{1}}\leftrightarrow-\boldsymbol{k_{1}}\right)\right]\nonumber \\
T^{3h} & = & 2M_{11}\left(k_{1}\right)M_{11}\left(k_{2}\right)M_{12}\left(k_{1},k_{2}\right)\left[B^{PT}\left(\boldsymbol{k_{1}},\boldsymbol{k_{2}},-\boldsymbol{k_{1}}-\boldsymbol{k_{2}}\right)+\left(\boldsymbol{k_{1}}\leftrightarrow-\boldsymbol{k_{1}}\right)\right]\nonumber \\
 & + & M_{11}\left(k_{1}\right)^{2}M_{22}\left(k_{2},k_{2}\right)P^{PT}\left(k_{1}\right)^{2}+\left(k_{1}\leftrightarrow k_{2}\right)\nonumber \\
 & + & 4M_{11}\left(k_{1}\right)M_{11}\left(k_{2}\right)M_{22}\left(k_{1},k_{2}\right)P^{PT}\left(k_{1}\right)P^{PT}\left(k_{2}\right)\nonumber \\
 & + & 2M_{12}\left(k_{1},k_{2}\right)\left[P^{PT}\left(\left|\boldsymbol{k_{1}}+\boldsymbol{k_{2}}\right|\right)+\left(\boldsymbol{k_{1}}\leftrightarrow-\boldsymbol{k_{1}}\right)\right]\left[M_{21}\left(k_{1}\right)M_{11}\left(k_{2}\right)P^{PT}\left(k_{2}\right)+\left(k_{1}\leftrightarrow k_{2}\right)\right]\nonumber \\
T^{4h} & = & M_{11}\left(k_{1}\right)^{2}M_{11}\left(k_{2}\right)^{2}T^{PT}\left(\boldsymbol{k_{1}},-\boldsymbol{k_{1}},\boldsymbol{k_{2}},-\boldsymbol{k_{2}}\right)\nonumber \\
 & + & 2M_{31}\left(k_{1}\right)M_{11}\left(k_{1}\right)M_{11}\left(k_{2}\right)^{2}P^{PT}\left(k_{1}\right)P^{PT}\left(k_{2}\right)^{2}+\left(k_{1}\leftrightarrow k_{2}\right)\nonumber \\
 & + & 2M_{21}\left(k_{1}\right)M_{11}\left(k_{1}\right)M_{11}\left(k_{2}\right)^{2}P^{PT}\left(k_{2}\right)\left[B^{PT}\left(\boldsymbol{k_{1}},\boldsymbol{k_{2}},-\boldsymbol{k_{1}}-\boldsymbol{k_{2}}\right)+\left(\boldsymbol{k_{1}}\leftrightarrow-\boldsymbol{k_{1}}\right)\right]+\left(k_{1}\leftrightarrow k_{2}\right)\nonumber \\
 & + & 2M_{21}\left(k_{1}\right)M_{11}\left(k_{2}\right)P^{PT}\left(k_{2}\right)\left[M_{21}\left(k_{1}\right)M_{11}\left(k_{2}\right)P^{PT}\left(k_{2}\right)+M_{11}\left(k_{1}\right)M_{21}\left(k_{2}\right)P^{PT}\left(k_{1}\right)\right]\nonumber \\
 &  & \hphantom{2M_{21}\left(k_{1}\right)M_{11}\left(k_{2}\right)P^{PT}\left(k_{2}\right)}\times\left[P^{PT}\left(\left|\boldsymbol{k_{1}}+\boldsymbol{k_{2}}\right|\right)+P^{PT}\left(\left|\boldsymbol{k_{1}}-\boldsymbol{k_{2}}\right|\right)\right]+\left(k_{1}\leftrightarrow k_{2}\right)
.\end{eqnarray}
This formula corrects the one of \cite{cooray2001powerspectrum},
which is incomplete for both the three-halo and four-halo terms.

A code implementation in Python computing this trispectrum in the
halo model is available upon request to the author.

\end{appendix}

\bibliographystyle{aa}
\bibliography{bibliography}

\begin{thebibliography}{34}
\expandafter\ifx\csname natexlab\endcsname\relax\def\natexlab#1{#1}\fi

\bibitem[{Benjamin {et~al.}(2012)Benjamin, Van~Waerbeke, Heymans, Kilbinger,
  Erben, Hildebrandt, Hoekstra, Kitching, Mellier, Miller, Rowe, Schrabback,
  Simpson, Coupon, Fu, Harnois-D{\'e}raps, Hudson, Kuijken, Semboloni, Vafaei,
  \& Velander}]{benjamin2012cfhtlens}
Benjamin, J., Van~Waerbeke, L., Heymans, C., {et~al.} 2012, arXiv:1212.3327
  [astro-ph], arXiv: 1212.3327

\bibitem[{Bernardeau {et~al.}(2002)Bernardeau, Colombi, Gaztanaga, \&
  Scoccimarro}]{bernardeau2002largescale}
Bernardeau, F., Colombi, S., Gaztanaga, E., \& Scoccimarro, R. 2002, Physics
  Reports, 367, 1, arXiv: astro-ph/0112551

\bibitem[{Bernardeau {et~al.}(2010)Bernardeau, Pitrou, \&
  Uzan}]{bernardeau2010cmbspectra}
Bernardeau, F., Pitrou, C., \& Uzan, J.-P. 2010, arXiv:1012.2652 [astro-ph,
  physics:gr-qc], arXiv: 1012.2652

\bibitem[{Bianchini {et~al.}(2015)Bianchini, Bielewicz, Lapi, Gonzalez-Nuevo,
  Baccigalupi, de~Zotti, Danese, Bourne, Cooray, Dunne, Dye, Eales, Ivison,
  Maddox, Negrello, Scott, Smith, \& Valiante}]{bianchini2015crosscorrelation}
Bianchini, F., Bielewicz, P., Lapi, A., {et~al.} 2015, The Astrophysical
  Journal, 802, 64, arXiv: 1410.4502

\bibitem[{Bleem {et~al.}(2012)Bleem, van Engelen, Holder, Aird, Armstrong,
  Ashby, Becker, Benson, Biesiadzinski, Brodwin, Busha, Carlstrom, Chang, Cho,
  Crawford, Crites, de~Haan, Desai, Dobbs, Dor{\'e}, Dudley, Geach, George,
  Gladders, Gonzalez, Halverson, Harrington, High, Holden, Holzapfel, Hoover,
  Hrubes, Joy, Keisler, Knox, Lee, Leitch, Lueker, Luong-Van, Marrone,
  Martinez-Manso, McMahon, Mehl, Meyer, Mohr, Montroy, Natoli, Padin, Plagge,
  Pryke, Reichardt, Rest, Ruhl, Saliwanchik, Sayre, Schaffer, Shaw, Shirokoff,
  Spieler, Stalder, Stanford, Staniszewski, Stark, Stern, Story, Vallinotto,
  Vanderlinde, Vieira, Wechsler, Williamson, \& Zahn}]{bleem2012ameasurement}
Bleem, L.~E., van Engelen, A., Holder, G.~P., {et~al.} 2012, The Astrophysical
  Journal, 753, L9, arXiv: 1203.4808

\bibitem[{Bryan \& Norman(1998)}]{bryan1998statistical}
Bryan, G.~L. \& Norman, M.~L. 1998, The Astrophysical Journal, 495, 80, arXiv:
  astro-ph/9710107

\bibitem[{Bullock {et~al.}(2001)Bullock, Kolatt, Sigad, Somerville, Kravtsov,
  Klypin, Primack, \& Dekel}]{bullock2001profiles}
Bullock, J.~S., Kolatt, T.~S., Sigad, Y., {et~al.} 2001, Monthly Notices of the
  Royal Astronomical Society, 321, 559, arXiv: astro-ph/9908159

\bibitem[{Collaboration {et~al.}(2015)Collaboration, Ade, Aghanim, Arnaud,
  Ashdown, Aumont, Baccigalupi, Banday, Barreiro, Bartlett, Bartolo, Battaner,
  Benabed, Beno{\^i}t, Benoit-L{\'e}vy, Bernard, Bersanelli, Bielewicz,
  Bonaldi, Bonavera, Bond, Borrill, Bouchet, Boulanger, Bucher, Burigana,
  Butler, Calabrese, Cardoso, Catalano, Challinor, Chamballu, Chiang,
  Christensen, Church, Clements, Colombi, Colombo, Combet, Couchot, Coulais,
  Crill, Curto, Cuttaia, Danese, Davies, Davis, de~Bernardis, de~Rosa,
  de~Zotti, Delabrouille, D{\'e}sert, Diego, Dole, Donzelli, Dor{\'e}, Douspis,
  Ducout, Dunkley, Dupac, Efstathiou, Elsner, En{\ss}lin, Eriksen, Fergusson,
  Finelli, Forni, Frailis, Fraisse, Franceschi, Frejsel, Galeotta, Galli,
  Ganga, Giard, Giraud-H{\'e}raud, Gjerl{\o}w, Gonz{\'a}lez-Nuevo, G{\'o}rski,
  Gratton, Gregorio, Gruppuso, Gudmundsson, Hansen, Hanson, Harrison,
  Henrot-Versill{\'e}, Hern{\'a}ndez-Monteagudo, Herranz, Hildebrandt, Hivon,
  Hobson, Holmes, Hornstrup, Hovest, Huffenberger, Hurier, Jaffe, Jaffe, Jones,
  Juvela, Keih{\"a}nen, Keskitalo, Kisner, Kneissl, Knoche, Kunz, Kurki-Suonio,
  Lagache, L{\"a}hteenm{\"a}ki, Lamarre, Lasenby, Lattanzi, Lawrence, Leonardi,
  Lesgourgues, Levrier, Lewis, Liguori, Lilje, Linden-V{\o}rnle,
  L{\'o}pez-Caniego, Lubin, Mac{\'i}as-P{\'e}rez, Maggio, Maino, Mandolesi,
  Mangilli, Martin, Mart{\'i}nez-Gonz{\'a}lez, Masi, Matarrese, Mazzotta,
  McGehee, Meinhold, Melchiorri, Mendes, Mennella, Migliaccio, Mitra,
  Miville-Desch{\^e}nes, Moneti, Montier, Morgante, Mortlock, Moss, Munshi,
  Murphy, Naselsky, Nati, Natoli, Netterfield, N{\o}rgaard-Nielsen, Noviello,
  Novikov, Novikov, Oxborrow, Paci, Pagano, Pajot, Paoletti, Pasian, Patanchon,
  Perdereau, Perotto, Perrotta, Pettorino, Piacentini, Piat, Pierpaoli,
  Pietrobon, Plaszczynski, Pointecouteau, Polenta, Popa, Pratt, Pr{\'e}zeau,
  Prunet, Puget, Rachen, Reach, Rebolo, Reinecke, Remazeilles, Renault, Renzi,
  Ristorcelli, Rocha, Rosset, Rossetti, Roudier, Rowan-Robinson,
  Rubi{\~n}o-Mart{\'i}n, Rusholme, Sandri, Santos, Savelainen, Savini, Scott,
  Seiffert, Shellard, Spencer, Stolyarov, Stompor, Sudiwala, Sunyaev, Sutton,
  Suur-Uski, Sygnet, Tauber, Terenzi, Toffolatti, Tomasi, Tristram, Tucci,
  Tuovinen, Valenziano, Valiviita, Van~Tent, Vielva, Villa, Wade, Wandelt,
  Wehus, White, Yvon, Zacchei, \& Zonca}]{planckcollaboration2015planck2015}
Collaboration, P., Ade, P. A.~R., Aghanim, N., {et~al.} 2015, arXiv:1502.01591
  [astro-ph], arXiv: 1502.01591

\bibitem[{Cooray \& Hu(2001)}]{cooray2001powerspectrum}
Cooray, A. \& Hu, W. 2001, The Astrophysical Journal, 554, 56, arXiv:
  astro-ph/0012087

\bibitem[{Cooray \& Sheth(2002)}]{cooray2002halomodels}
Cooray, A. \& Sheth, R. 2002, Physics Reports, 372, 1, arXiv: astro-ph/0206508

\bibitem[{Erben {et~al.}(2013)Erben, Hildebrandt, Miller, van Waerbeke,
  Heymans, Hoekstra, Kitching, Mellier, Benjamin, Blake, Bonnett, Cordes,
  Coupon, Fu, Gavazzi, Gillis, Grocutt, Gwyn, Holhjem, Hudson, Kilbinger,
  Kuijken, Milkeraitis, Rowe, Schrabback, Semboloni, Simon, Smit, Toader,
  Vafaei, van Uitert, \& Velander}]{erben2013cfhtlens}
Erben, T., Hildebrandt, H., Miller, L., {et~al.} 2013, Monthly Notices of the
  Royal Astronomical Society, 433, 2545

\bibitem[{Foreman-Mackey {et~al.}(2013)Foreman-Mackey, Hogg, Lang, \&
  Goodman}]{foreman-mackey2013emceethe}
Foreman-Mackey, D., Hogg, D.~W., Lang, D., \& Goodman, J. 2013, Publications of
  the Astronomical Society of the Pacific, 125, 306, arXiv: 1202.3665

\bibitem[{Geach {et~al.}(2013)Geach, Hickox, Bleem, Brodwin, Holder, Aird,
  Benson, Bhattacharya, Carlstrom, Chang, Cho, Crawford, Crites, de~Haan,
  Dobbs, Dudley, George, Hainline, Halverson, Holzapfel, Hoover, Hou, Hrubes,
  Keisler, Knox, Lee, Leitch, Lueker, Luong-Van, Marrone, McMahon, Mehl, Meyer,
  Millea, Mohr, Montroy, Myers, Padin, Plagge, Pryke, Reichardt, Ruhl, Sayre,
  Schaffer, Shaw, Shirokoff, Spieler, Staniszewski, Stark, Story, van Engelen,
  Vanderlinde, Vieira, Williamson, \& Zahn}]{geach2013adirect}
Geach, J.~E., Hickox, R.~C., Bleem, L.~E., {et~al.} 2013, The Astrophysical
  Journal, 776, L41

\bibitem[{Goodman \& Weare(2010)}]{goodman2010ensemble}
Goodman, J. \& Weare, J. 2010, Communications in Applied Mathematics and
  Computational Science, 5, 65

\bibitem[{Goroff {et~al.}(1986)Goroff, Grinstein, Rey, \&
  Wise}]{goroff1986coupling}
Goroff, M.~H., Grinstein, B., Rey, S.-J., \& Wise, M.~B. 1986, The
  Astrophysical Journal, 311, 6

\bibitem[{G{\'o}rski {et~al.}(2005)G{\'o}rski, Hivon, Banday, Wandelt, Hansen,
  Reinecke, \& Bartelmann}]{gorski2005healpix}
G{\'o}rski, K.~M., Hivon, E., Banday, A.~J., {et~al.} 2005, The Astrophysical
  Journal, 622, 759

\bibitem[{Gunn \& Gott(1972)}]{gunn1972onthe}
Gunn, J.~E. \& Gott, III, J.~R. 1972, The Astrophysical Journal, 176, 1

\bibitem[{Heymans {et~al.}(2012)Heymans, Van~Waerbeke, Miller, Erben,
  Hildebrandt, Hoekstra, Kitching, Mellier, Simon, Bonnett, Coupon, Fu,
  Harnois~D{\'e}raps, Hudson, Kilbinger, Kuijken, Rowe, Schrabback, Semboloni,
  van Uitert, Vafaei, \& Velander}]{heymans2012cfhtlens}
Heymans, C., Van~Waerbeke, L., Miller, L., {et~al.} 2012, Monthly Notices of
  the Royal Astronomical Society, 427, 146

\bibitem[{Hirata {et~al.}(2008)Hirata, Ho, Padmanabhan, Seljak, \&
  Bahcall}]{hirata2008correlation}
Hirata, C.~M., Ho, S., Padmanabhan, N., Seljak, U., \& Bahcall, N. 2008,
  Physical Review D, 78, arXiv: 0801.0644

\bibitem[{Lewis \& Bridle(2002)}]{lewis2002cosmological}
Lewis, A. \& Bridle, S. 2002, Physical Review D, 66, 103511

\bibitem[{LoVerde \& Afshordi(2008)}]{loverde2008extended}
LoVerde, M. \& Afshordi, N. 2008, Physical Review D, 78, arXiv: 0809.5112

\bibitem[{Mo \& White(1995)}]{mo1995ananalytic}
Mo, H.~J. \& White, S. D.~M. 1995, arXiv:astro-ph/9512127, arXiv:
  astro-ph/9512127

\bibitem[{Moessner {et~al.}(1998)Moessner, Jain, \&
  Villumsen}]{moessner1998theeffect}
Moessner, R., Jain, B., \& Villumsen, J.~V. 1998, Monthly Notices of the Royal
  Astronomical Society, 294, 291, arXiv: astro-ph/9708271

\bibitem[{Navarro {et~al.}(1997)Navarro, Frenk, \&
  White}]{navarro1997auniversal}
Navarro, J.~F., Frenk, C.~S., \& White, S. D.~M. 1997, The Astrophysical
  Journal, 490, 493, arXiv: astro-ph/9611107

\bibitem[{Okamoto \& Hu(2003)}]{okamoto2003cmblensing}
Okamoto, T. \& Hu, W. 2003, Physical Review D, 67, arXiv: astro-ph/0301031

\bibitem[{Omori \& Holder(2015)}]{omori2015crosscorrelation}
Omori, Y. \& Holder, G. 2015, arXiv:1502.03405 [astro-ph], arXiv: 1502.03405

\bibitem[{Peter \& Uzan(2013)}]{peter2013primordial}
Peter, P. \& Uzan, J.-P. 2013, Primordial cosmology, 1st edn., Oxford graduate
  texts (Oxford: Oxford Univ. Press)

\bibitem[{{Planck Collaboration} {et~al.}(2014){Planck Collaboration}, Ade,
  Aghanim, Armitage-Caplan, Arnaud, Ashdown, Atrio-Barandela, Aumont,
  Baccigalupi, Banday, Barreiro, Bartlett, Basak, Battaner, Benabed,
  Beno{\^i}t, Benoit-L{\'e}vy, Bernard, Bersanelli, Bielewicz, Bobin, Bock,
  Bonaldi, Bonavera, Bond, Borrill, Bouchet, Bridges, Bucher, Burigana, Butler,
  Cardoso, Catalano, Challinor, Chamballu, Chiang, Chiang, Christensen, Church,
  Clements, Colombi, Colombo, Couchot, Coulais, Crill, Curto, Cuttaia, Danese,
  Davies, Davis, de~Bernardis, de~Rosa, de~Zotti, D{\'e}chelette, Delabrouille,
  Delouis, D{\'e}sert, Dickinson, Diego, Dole, Donzelli, Dor{\'e}, Douspis,
  Dunkley, Dupac, Efstathiou, En{\ss}lin, Eriksen, Finelli, Forni, Frailis,
  Franceschi, Galeotta, Ganga, Giard, Giardino, Giraud-H{\'e}raud,
  Gonz{\'a}lez-Nuevo, G{\'o}rski, Gratton, Gregorio, Gruppuso, Gudmundsson,
  Hansen, Hanson, Harrison, Henrot-Versill{\'e}, Hern{\'a}ndez-Monteagudo,
  Herranz, Hildebrandt, Hivon, Ho, Hobson, Holmes, Hornstrup, Hovest,
  Huffenberger, Jaffe, Jaffe, Jones, Juvela, Keih{\"a}nen, Keskitalo, Kisner,
  Kneissl, Knoche, Knox, Kunz, Kurki-Suonio, Lagache, L{\"a}hteenm{\"a}ki,
  Lamarre, Lasenby, Laureijs, Lavabre, Lawrence, Leahy, Leonardi,
  Le{\'o}n-Tavares, Lesgourgues, Lewis, Liguori, Lilje, Linden-V{\o}rnle,
  L{\'o}pez-Caniego, Lubin, Mac{\'i}as-P{\'e}rez, Maffei, Maino, Mandolesi,
  Mangilli, Maris, Marshall, Martin, Mart{\'i}nez-Gonz{\'a}lez, Masi, Massardi,
  Matarrese, Matthai, Mazzotta, Melchiorri, Mendes, Mennella, Migliaccio,
  Mitra, Miville-Desch{\^e}nes, Moneti, Montier, Morgante, Mortlock, Moss,
  Munshi, Murphy, Naselsky, Nati, Natoli, Netterfield, N{\o}rgaard-Nielsen,
  Noviello, Novikov, Novikov, Osborne, Oxborrow, Paci, Pagano, Pajot, Paoletti,
  Partridge, Pasian, Patanchon, Perdereau, Perotto, Perrotta, Piacentini, Piat,
  Pierpaoli, Pietrobon, Plaszczynski, Pointecouteau, Polenta, Ponthieu, Popa,
  Poutanen, Pratt, Pr{\'e}zeau, Prunet, Puget, Pullen, Rachen, Rebolo,
  Reinecke, Remazeilles, Renault, Ricciardi, Riller, Ristorcelli, Rocha,
  Rosset, Roudier, Rowan-Robinson, Rubi{\~n}o-Mart{\'i}n, Rusholme, Sandri,
  Santos, Savini, Scott, Seiffert, Shellard, Smith, Spencer, Starck, Stolyarov,
  Stompor, Sudiwala, Sunyaev, Sureau, Sutton, Suur-Uski, Sygnet, Tauber,
  Tavagnacco, Terenzi, Toffolatti, Tomasi, Tristram, Tucci, Tuovinen, Umana,
  Valenziano, Valiviita, Van~Tent, Vielva, Villa, Vittorio, Wade, Wandelt,
  White, White, Yvon, Zacchei, \& Zonca}]{planckcollaboration2014textitplanck}
{Planck Collaboration}, Ade, P. A.~R., Aghanim, N., {et~al.} 2014, Astronomy \&
  Astrophysics, 571, A17

\bibitem[{Press \& Schechter(1974)}]{press1974formation}
Press, W.~H. \& Schechter, P. 1974, The Astrophysical Journal, 187, 425

\bibitem[{Scherrer \& Bertschinger(1991)}]{scherrer1991statistics}
Scherrer, R.~J. \& Bertschinger, E. 1991, The Astrophysical Journal, 381, 349

\bibitem[{Sherwin {et~al.}(2012)Sherwin, Das, Hajian, Addison, Bond, Crichton,
  Devlin, Dunkley, Gralla, Halpern, Hill, Hincks, Hughes, Huffenberger, Hlozek,
  Kosowsky, Louis, Marriage, Marsden, Menanteau, Moodley, Niemack, Page, Reese,
  Sehgal, Sievers, Sifon, Spergel, Staggs, Switzer, \&
  Wollack}]{sherwin2012theatacama}
Sherwin, B.~D., Das, S., Hajian, A., {et~al.} 2012, Physical Review D, 86,
  arXiv: 1207.4543

\bibitem[{Sheth \& Tormen(1999)}]{sheth1999largescale}
Sheth, R.~K. \& Tormen, G. 1999, Monthly Notices of the Royal Astronomical
  Society, 308, 119

\bibitem[{Smith {et~al.}(2007)Smith, Zahn, \& Dore}]{smith2007detection}
Smith, K.~M., Zahn, O., \& Dore, O. 2007, Physical Review D, 76, arXiv:
  0705.3980

\bibitem[{Takada \& Hu(2013)}]{takada2013powerspectrum}
Takada, M. \& Hu, W. 2013, Physical Review D, 87, arXiv: 1302.6994

\end{thebibliography}

\end{document}